\documentclass{emulateapj}

\usepackage{natbib}

\newcommand{\msun}{M$_{\sun}$}

\newcommand{\ldl}{$\lambda/{\Delta}{\lambda}$}
\newcommand{\teff}{T$_{eff}$}

\newcommand{\meth}{CH$_4$}

\newcommand{\wat}{H$_2$O}

\newcommand{\lha}{$\log_{10}{L_{H\alpha}/L_{bol}}$}
\newcommand{\kms}{km~s$^{-1}$}

\newcommand{\masyr}{mas~yr$^{-1}$}
\newcommand{\name}{SDSS~J000649.16$-$085246.3}
\newcommand{\namesh}{SDSS~J0006$-$0852}

\slugcomment{Submitted to ApJ on 30 April 2012, Accepted 1 August 2012}

\begin{document}

\title{Discovery of a Very Low Mass Triple with Late-M and T Dwarf Components: LP~704-48/SDSS~J0006$-$0852AB\footnote{Portions of the data presented herein were obtained at the W.M. Keck Observatory, which is operated as a scientific partnership among the California Institute of Technology, the University of California and the National Aeronautics and Space Administration. The Observatory was made possible by the generous financial support of the W.M. Keck Foundation}}

\author{Adam J. Burgasser\altaffilmark{1,2},
Christopher Luk\altaffilmark{1},
Saurav Dhital\altaffilmark{3,4},
Daniella Bardalez Gagliuffi\altaffilmark{1},
Christine P.\ Nicholls\altaffilmark{1},
L.\ Prato\altaffilmark{5},
Andrew A. West\altaffilmark{4}
and S\'{e}bastien L\'{e}pine\altaffilmark{6}}

\altaffiltext{1}{Center for Astrophysics and Space Science, University of California San Diego, La Jolla, CA, 92093, USA; aburgasser@ucsd.edu}
\altaffiltext{2}{Visiting Astronomer at the Infrared Telescope Facility, which is operated by the University of Hawaii under Cooperative Agreement no. NNX-08AE38A with the National Aeronautics and Space Administration, Science Mission Directorate, Planetary Astronomy Program.}
\altaffiltext{3}{Department of Physics \& Astronomy, Vanderbilt University, Nashville, TN, 37235, USA}
\altaffiltext{4}{Department of Astronomy, Boston University, 725 Commonwealth Ave Boston, MA 02215 USA}
\altaffiltext{5}{Lowell Observatory, Flagstaff, AZ 86001 USA}
\altaffiltext{6}{Department of Astrophysics, Division of Physical Sciences,
American Museum of Natural History, Central Park West at 79th
Street, New York, NY 10024, USA}

\begin{abstract}
We report the identification of the M9 dwarf SDSS~J000649.16$-$085246.3 as a spectral binary and radial velocity variable 
with components straddling the hydrogen burning mass limit.  
Low-resolution near-infrared spectroscopy reveals spectral features indicative of a T dwarf companion, and spectral template fitting yields component types of M8.5$\pm$0.5 and T5$\pm$1.
High-resolution near-infrared spectroscopy with Keck/NIRSPEC reveals pronounced radial velocity variations with a semi-amplitude of 8.2$\pm$0.4~{\kms}.  From these
we determine an orbital period of 147.6$\pm$1.5~days and eccentricity of 0.10$\pm$0.07, making SDSS~J0006$-$0852AB the third tightest very low mass binary known. This system is also found to have a common
proper motion companion, the inactive M7 dwarf LP~704-48, at a projected separation of 820$\pm$120~AU.
The lack of H$\alpha$ emission in both M dwarf components indicates that this system is relatively old, as confirmed by evolutionary model analysis of the tight binary.
LP~704-48/SDSS~J0006$-$0852AB is the lowest-mass confirmed triple identified to date, 
and one of only seven candidate and confirmed triples with total masses below 0.3~{\msun} currently known.
We show that current star and brown dwarf formation models cannot produce triple systems like LP~704-48/SDSS~J0006$-$0852AB, and we rule out Kozai-Lidov perturbations and tidal circularization as a viable mechanism to shrink the inner orbit.
The similarities between this system and the recently uncovered low-mass eclipsing triples 
NLTT~41135AB/41136 and LHS~6343ABC suggest that substellar tertiaries may be common in wide M dwarf pairs.  
\end{abstract}


\section{Introduction}

Our current theoretical and observational understanding of star formation holds that most stars form in multiple systems, with the frequency, mass ratio distribution and period distribution of multiples varying as a function of mass and possibly formation environment (e.g., \citealt{2004MNRAS.351..617D,2007ApJ...671.2074A,2007prpl.conf..427B,2009MNRAS.392..590B,2012MNRAS.419.3115B,2011ApJ...731....8K}).  Dynamics play an important role in multiple formation, as both the fragmentation of collapsing cloud cores \citep{2001ApJ...551L.167B} and massive circumstellar disks \citep{2009MNRAS.392..413S,2011MNRAS.413.1787S} give rise to small N groups that either dissolve or stabilize into hierarchical systems \citep{2003A&A...400.1031S}.  Numerical calculations show that both fragmentation and dynamical evolution of stellar multiples are sensitive to the initial cloud properties---geometry, density, turbulence spectrum, and radiative and external feedback---so multiplicity statistics provide important tests for star formation theory, on par with the initial mass function, stellar mass segregation and velocity distributions.

The mass dependence of multiplicity among stars is well established, with observed frequencies ranging from $>$80\% for OBA dwarfs \citep{2002A&A...382...92S,2005A&A...430..137K,2007ApJ...670..747K} to $\sim$30\% for systems with M dwarf primaries \citep{1992ApJ...396..178F,1997AJ....113.2246R,2004ASPC..318..166D}.  The frequency of multiples appears to drop even further in the very low mass (VLM; M $\lesssim$ 0.1~{\msun}) stellar and substellar regimes.  Resolved imaging studies of field and cluster VLM dwarfs, based on ground-based adaptive optics (AO) and Hubble Space Telescope (HST) observations, find relatively consistent (volume-limited) binary frequencies of 10--20\% (e.g., \citealt{2003AJ....126.1526B,2003ApJ...586..512B,2006ApJS..166..585B,2003ApJ...587..407C}).  However, estimates for the frequency of tightly-bound spectroscopic VLM binaries (roughly 10\% of currently known systems) span a broad range of 1--25\% \citep{2005MNRAS.362L..45M,2006AJ....132..663B,2006MNRAS.372.1879K,2008A&A...492..545J,2010ApJ...723..684B,2012ApJ...744..119C}.  As such, there remains significant uncertainty in the overall VLM binary frequency, which has a direct impact on our understanding of how brown dwarfs form in the first place (e.g., \citealt{2003MNRAS.342..926D}). There is also evidence that the orbital characteristics of multiples are mass-dependent, with VLM binaries being on average tighter ($\langle{a}\rangle \approx$ 7~AU versus $\approx$ 30~AU) and more frequently composed of equal-mass components ($\langle{q}\rangle \equiv {\rm M}_2/{\rm M}_1 \approx 1$ versus $\approx$ 0.3) than their solar-mass counterparts \citep{2007ApJ...668..492A}.  Again, these trends may be skewed by biases inherent to the known sample of VLM binaries, identified largely as imaged pairs.  A complete understanding of VLM multiplicity statistics therefore requires more robust constraints on the frequency and characteristics of short-period and low-$q$ multiples \citep{2007prpl.conf..427B}.

An alternative method for identifying VLM multiples that avoids separation biases (both intrinsic and projected) is through spectral binaries, systems with components of different spectral types whose combined-light spectra exhibit distinct peculiarities.  While this technique is more commonly associated with white dwarf-M dwarf pairs (e.g., \citealt{2006AJ....131.1674S}), spectral binaries containing late M or L dwarf primaries and T dwarf secondaries have also been recognized due to their unique and highly structured near-infrared spectra  \citep{2004ApJ...604L..61C,2007AJ....134.1330B,2008ApJ...681..579B,2008arXiv0811.0556S,2010ApJ...710.1142B,2011ApJ...739...49B,2010AJ....140..110G,2011ApJ...732...56G}.  Because blended light systems can be identified at any separation, M/L+T spectral binaries can probe the very closest separations that are inaccessible to direct imaging studies ($a \lesssim$ 50~mas).  
An illustrative case is the M8.5 + T5 binary 2MASS~J03202839$-$0446358 (hereafter 2MASS~J0320$-$0446), a system independently identified as both a spectral binary \citep{2008ApJ...681..579B} and a radial velocity (RV) variable with a period of 0.68~yr and primary semi-major axis $a_1\sin{i}$ = 0.157$\pm$0.003~AU  \citep{2008ApJ...678L.125B,2010ApJ...723..684B}. With a projected separation of $\approx$17~mas, this system is unresolvable with current imaging and interferometric technology.
Moreover, the significant difference in component magnitudes ($\Delta{K}$ = 4.3$\pm$0.6) makes this system
a potential low $q$ pair.  
The independent constraints provided by the component classifications and RV orbit yield robust limits on the age ($\gtrsim$2~Gyr) and orbital inclination ($\gtrsim 53\degr$) of this system \citep{2009AJ....137.4621B}.  

In an effort to confirm and characterize spectral binaries identified in low-resolution near-infrared spectroscopy, we have identified a new VLM system exhibiting significant RV variability.  The source, {\name} (hereafter {\namesh}; \citealt{2008AJ....135..785W}) is an optically-classified M9 which appears to be both a short-period (0.4~yr) binary with M8.5 and T5$\pm$1 components, and a co-moving wide companion to the inactive M7 dwarf LP~704-48.  Together, this system comprises the lowest-mass triple confirmed to date.
In Section~2 we describe our optical and near-infrared imaging and spectroscopic observations of this system using 
the 1m Lick Observatory Nickel Direct Imaging Camera, 
the 3m NASA Infrared Telescope Facility (IRTF) SpeX spectrograph \citep{2003PASP..115..362R},
the 4m Kitt Peak National Observatory (KPNO) RC spectrograph, 
and the 10m Keck II NIRSPEC spectrograph \citep{1998SPIE.3354..566M}.
In Section~3 we examine the properties of the {\namesh} and LP~704-48 pair, assessing their common proper motion, distance and probability of chance alignment; as well as the age and metallicity of
both components as derived from spectroscopic and kinematic indicators.
In Section~4 we analyze the low-resolution near-infrared spectrum of {\namesh} to infer 
the presence of, and characterize, its T dwarf companion.
In Section~5 we analyze our RV measurements which allow us to extract both the orbital properties (including constraints on the orbital inclination) and verify the relatively old age for this system.  
In Section~6 we discuss LP~704-48/{\namesh}AB in the context of other 
hierarchical low-mass triples, and examine whether current star formation theories
or three-body interactions can create such systems. 
Our results are summarized in Section~7.

\section{Observations}

\subsection{Optical Imaging}

\begin{figure}
\epsscale{0.7}
\plotone{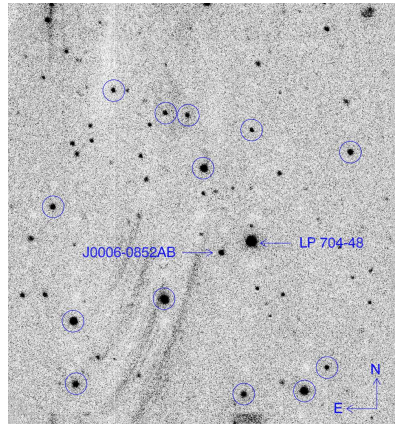}
\caption{Lick/Nickel image of the LP~704-48/{\namesh} field showing the location of the two sources (arrows and labels) and field sources used for calibrating the astrometric reference frame (circles).  The field shown is approximately 5$\arcmin\times$5$\arcmin$ and oriented North up and East to the left.   Streaking along the bottom of the image is due to reflected lunar light. 
\label{fig:nickelim}}
\end{figure}

The {\namesh} field was observed with the Nickel 1m CCD camera on 2011 October 28 (UT) in clear conditions with 1$\farcs$5 seeing.  The chip was binned 2$\times$2 for a pixel scale of 0$\farcs$37 pixel$^{-1}$.  Six dithered exposures of 150~s each were obtained using the $I$-band filter over an airmass range of 1.45--1.49.  Dome flat and dark exposures were obtained at the end of the night for pixel calibration.  Data were reduced using custom IDL\footnote{Interactive Data Language.} scripts that removed the bias voltage, median-combined and normalized the flat field frames; divided the science data by the flat; masked bad pixels; generated a median sky frame that was subtracted from the science data; and registered and median-stacked the science frames to a single final image, shown in Figure~\ref{fig:nickelim}.  

\begin{figure}
\epsscale{0.9}
\plotone{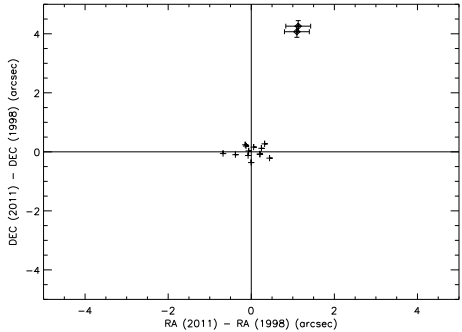}
\caption{Offsets in Right Ascension and declination between 2MASS and Nickel imaging epochs for field stars (plus symbols) and LP~704-48/{\namesh} (diamonds).
The error bars on the last two data points reflect the scatter in offsets among the field stars. 
\label{fig:nickelpm}}
\end{figure}

Pixel positions for sources in the final stacked frame were measured using a centroiding algorithm routine attached to the {\em atv} program \citep{2001ASPC..238..385B}.  Using astrometry of 13 field stars drawn from the Two Micron All Sky Survey (2MASS; \citealt{2006AJ....131.1163S}), we determined a linear astrometric transformation frame with residuals of 0$\farcs$3 in Right Ascension and 0$\farcs$2 and declination; i.e., $\sim$20\% of the seeing radius.  This transformation was used to compute coordinates for LP~704-48 and {\namesh}, listed in Table~\ref{tab:astrometry}. 
Figure~\ref{fig:nickelpm} displays the 2MASS to Nickel offsets for all of the sources in the field.  The large and similar proper motions for LP~704-48 and {\namesh} are evident, distinct from the field by $>20\sigma$ in declination.

\begin{deluxetable}{lcccc}
\tabletypesize{\scriptsize}
%\rotate
\tablecaption{Astrometry for {\namesh} and LP~704-48 \label{tab:astrometry}}
\tablewidth{0pt}
\tablehead{
\colhead{Source} &
\colhead{Epoch} &
\colhead{$\alpha$} &
\colhead{$\delta$} &
\colhead{Uncertainties} \\
 & \colhead{(UT)} &  \colhead{(J2000)} &  \colhead{(J2000)} & \colhead{($\arcsec$)} \\
}
\startdata
\tableline
\multicolumn{5}{c}{LP~704-48} \\
\tableline
POSS~I $R$ & 1954.669 & 00 06 47.66 & $-$08 52 21.17 & 0.1, 0.1\tablenotemark{a} \\
2MASS  & 1998.767 & 00 06 47.46 & $-$08 52 35.00 & 0.11, 0.09 \\
ESO $R$ & 1999.604 & 00 06 47.45  & $-$08 52 35.31 & 0.1, 0.1\tablenotemark{a} \\
UKST $I_N$ & 2000.723 & 00 06 47.45 & $-$08 52 35.66 & 0.1, 0.1\tablenotemark{a} \\
SDSS & 2000.740 & 00 06 47.45 & $-$08 52 35.60 & 0.06, 0.04 \\
Nickel & 2011.820 & 00 06 47.39 & $-$08 52 39.16 & 0.3, 0.2 \\
\tableline
\multicolumn{5}{c}{\namesh} \\
\tableline
2MASS  & 1998.767 & 00 06 49.16 & $-$08 52 45.70 & 0.11, 0.09 \\
ESO $R$ & 1999.604 & 00 06 49.15 & $-$08 52 45.53 & 0.1, 0.1\tablenotemark{a} \\
UKST $I_N$ & 2000.723 & 00 06 49.14  & $-$08 52 46.42 & 0.1, 0.1\tablenotemark{a} \\
SDSS & 2000.740 & 00 06 49.16 & $-$08 52 46.30 & 0.07, 0.06 \\
Nickel & 2011.820 & 00 06 49.09 & $-$08 52 49.96 & 0.3, 0.2 \\
\enddata
\tablenotetext{a}{Estimated.}
\tablerefs{POSS~1 $R$, ESO~$R$ and UKST~$I_N$ astrometry are from the SuperCosmos Sky Survey \citep{2001MNRAS.326.1279H,2001MNRAS.326.1295H,2001MNRAS.326.1315H}; 2MASS astrometry are from the All-Sky Data Release Point Source Catalog \citep{2006AJ....131.1163S}; SDSS astrometry are from Data Release 7 \citep{2009ApJS..182..543A}.}
\end{deluxetable}

\subsection{Low Resolution Optical Spectroscopy}

\begin{figure}
\epsscale{0.8}
\plotone{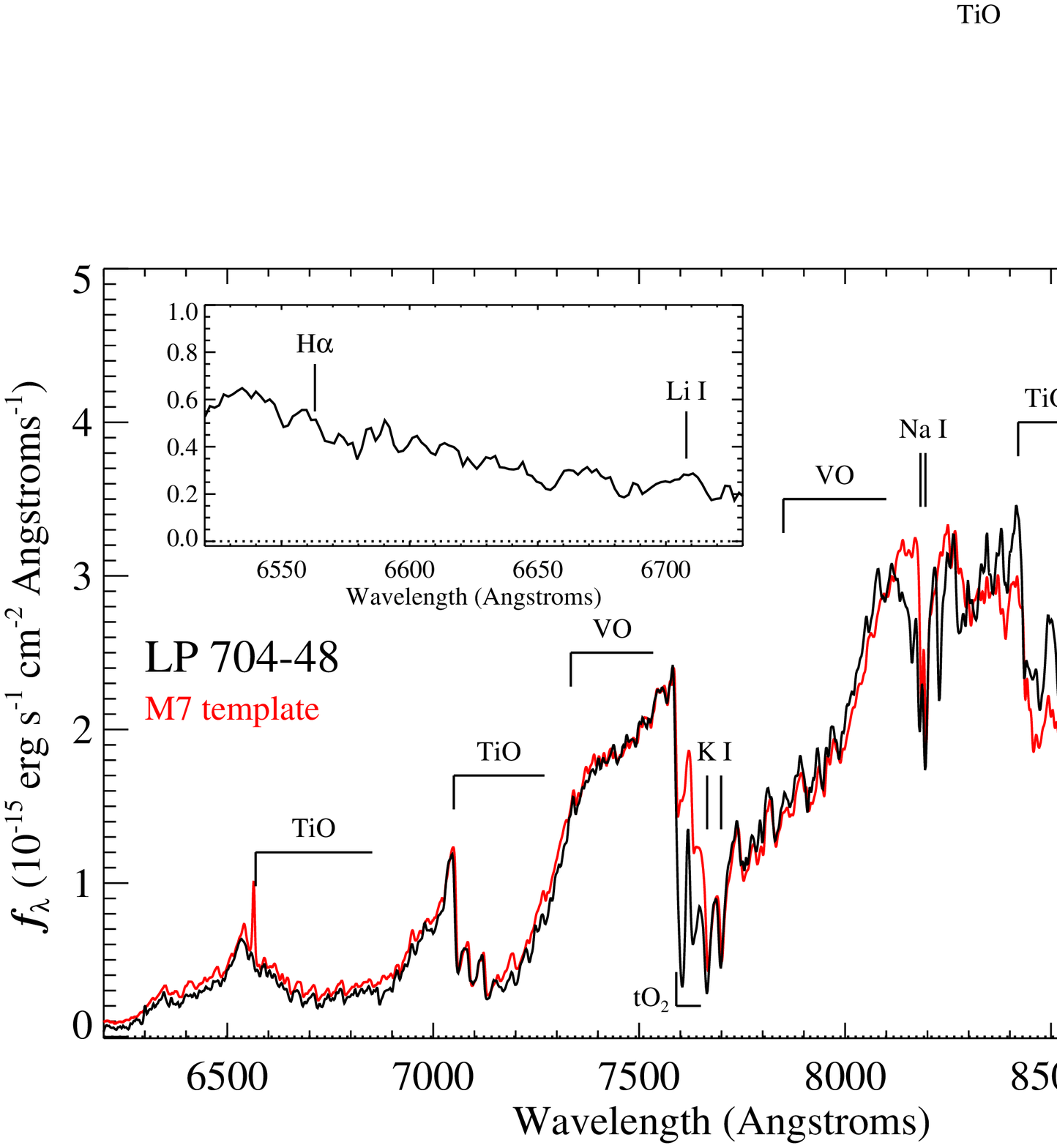}
\plotone{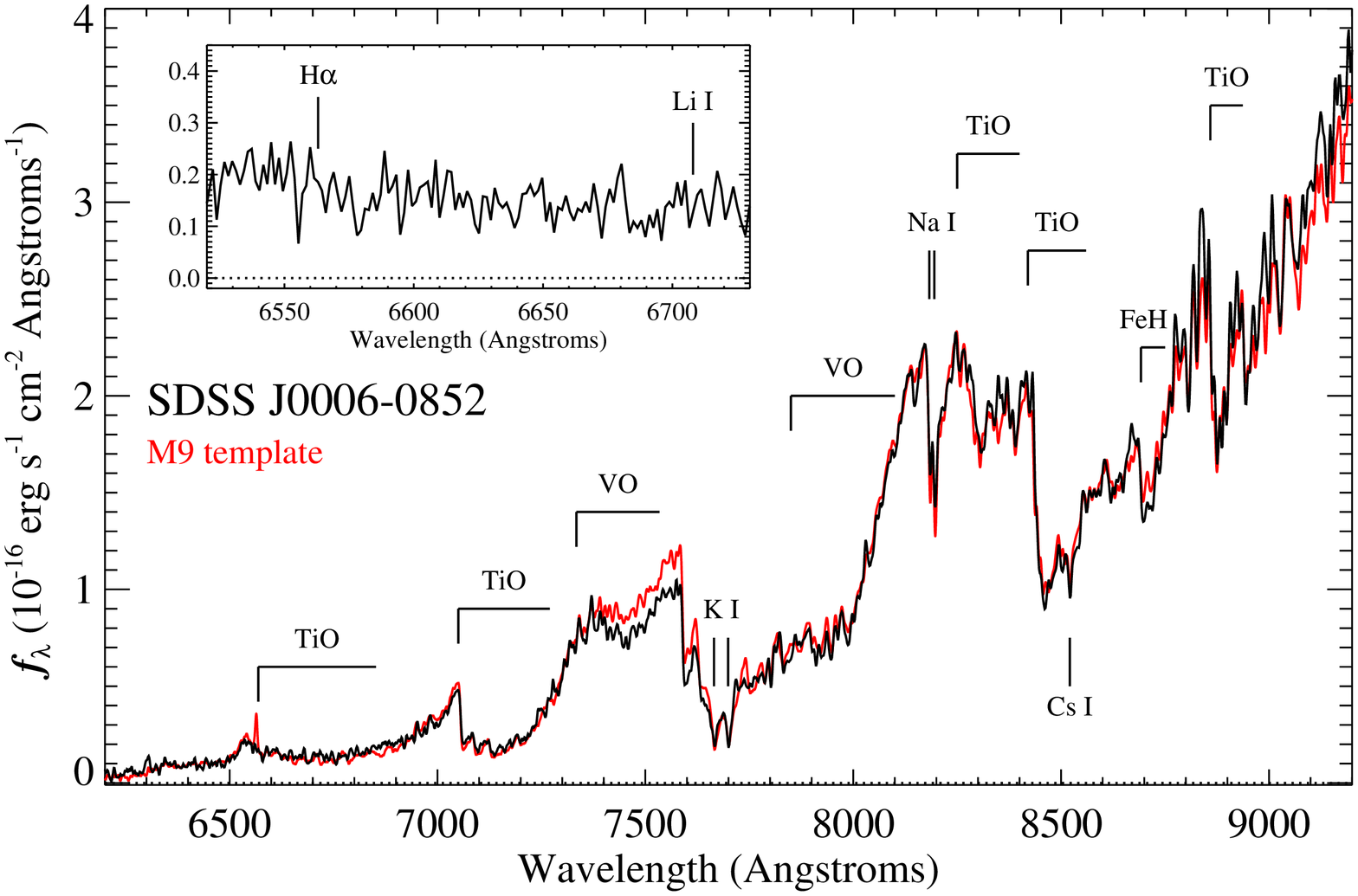}
\caption{Red optical spectra of LP~704-48 (top)  and {\namesh} (bottom, from SDSS) compared to M7 and M9  SDSS spectral templates from \citet{2007AJ....133..531B}. All spectra shown are smoothed to a common resolution of {\ldl} = 1200.  Data for the two dwarfs are scaled to their apparent SDSS $i$ magnitudes, while the templates are scaled to overlap in the 7400--7500~{\AA} (LP~704-48) and 8550--8600~{\AA} ({\namesh}) regions.  Primary atomic and molecular absorption features are labeled.  The inset boxes shows close-ups of the 6520--6730~{\AA} region, showing the absence of both H$\alpha$ emission and Li~I  absorption in both sources.
\label{fig:optspec}}
\end{figure}

The optical spectrum of {\namesh} from the Sloan Digital Sky Survey (SDSS; \citealt{2000AJ....120.1579Y}) is shown in Figure~\ref{fig:optspec} compared to an M9 SDSS spectral template from \citet{2007AJ....133..531B}.  The similarity of these spectra confirms the M9 classification from \citet{2008AJ....135..785W}, although we note a slightly flatter 7500~{\AA} VO band and weaker Na~I lines at 8200~{\AA} for {\namesh}.  Neither Li~I absorption (EW $<$ 0.3~{\AA}) nor H$\alpha$ emission ($f_{H\alpha} < 4\times10^{-18}$~erg~s$^{-1}$~cm$^{-2}$) are detected in these data.
Upper limits on {\lha} were calculated by finding the H$\alpha$ equivalent width corresponding to the 1$\sigma$ flux uncertainty, using the $\chi$ factor \citep{2004PASP..116.1105W,2008PASP..120.1161W} to convert from EW to L$_{\rm{H}\alpha}$/L$_{\rm{bol}}$ and to propagate uncertainties.
We find  {\lha} $<$ $-$5.7 for {\namesh}.

The nearby proper motion star LP~704-48 (aka 2MASS~J00064746$-$0852350; \citealt{1980nltt.bookQ....L})
was classified M6 by \citet{2002AJ....123.2828C} based on low-resolution optical
spectroscopy.
We obtained a new optical spectrum of LP~704-48 on 2011 Nov 28 (UT) using the RC spectrograph on the KPNO 4m telescope. Conditions were good, with seeing $\sim$1$\arcsec$ and excellent transparency. We used a 1$\farcs$5 slit aligned to the parallactic angle, a 600 lines~mm$^{-1}$ grating blazed at 7500~{\AA} (BL~420), and an order-blocking filter (OG~530) to obtain 5800--9100~{\AA} spectra at a resolution of $\sim$2000 and dispersion of 1.74~{\AA}~pixel$^{-1}$. A single 1200~s exposure was obtained at an airmass of 1.32. 
We observed the sdO flux standard HZ~4 for flux calibration, and obtained a set of quartz flats, bias frames, and arclamp exposures for pixel response and wavelength calibration.  Data were reduced using standard routines in IRAF\footnote{Image Reduction and Analysis Facility \citep{1986SPIE..627..733T}.}.

\begin{deluxetable*}{lccccccl}
\tablecaption{Optical Spectral Indices for LP~704-48 and {\namesh} \label{tab:optindices}}
\tabletypesize{\scriptsize}
%\rotate
\tablewidth{0pt}
\tablehead{
\colhead{Index} &
\multicolumn{2}{c}{LP~704-48} & 
\multicolumn{2}{c}{LP~704-48\tablenotemark{a}} & 
\multicolumn{2}{c}{\namesh} & 
\colhead{Ref}  \\
}
\startdata
TiO5 & 0.208$\pm$0.005 & & 0.213$\pm$0.010 & & 0.33$\pm$0.03 & & 1 \\
CaH2 & 0.256$\pm$0.004 & & 0.255$\pm$0.007 & & 0.35$\pm$0.03 & & 1 \\
CaH3 & 0.558$\pm$0.007 & & 0.569$\pm$0.013 & & 0.61$\pm$0.03 & & 1 \\
VO1 & 0.836$\pm$0.004 & (M6.5) & 0.826$\pm$0.008 & (M7) & 0.768$\pm$0.014 & (M9) & 2,3 \\
VO2 & 0.582$\pm$0.003 & (M6.5) & 0.610$\pm$0.006 & (M6) & 0.340$\pm$0.007 & (M9) & 3 \\
TiO7 & 0.707$\pm$0.006 & (M6) & 0.723$\pm$0.009 & (M6) & 0.504$\pm$0.014 & (M8) & 3 \\
Color-M & 4.85$\pm$0.03 & (M7) & 5.23$\pm$0.06 & (M7) & 9.41$\pm$0.15 & (M9.5) & 3 \\
$\zeta$ & 1.002$\pm$0.010 & & 1.004$\pm$0.014 & & 0.96$\pm$0.06 & & 4 \\
\enddata
\tablenotetext{a}{Data from \citet{2002AJ....123.2828C}.}
\tablenotetext{b}{Classifications based on the relations quantified in \citet{2003AJ....125.1598L}.}
\tablerefs{(1) \citet{1995AJ....110.1838R}; (2) \citet{2002AJ....123.3409H}; (3)  \citet{2003AJ....125.1598L}; (4) \citet{2007ApJ...669.1235L}.} 
\end{deluxetable*}

The reduced spectrum for LP~704-48 
is shown in Figure~\ref{fig:optspec}, compared to an M7 SDSS template from 
 \citet{2007AJ....133..531B}.  Both our data and those of \citet{2002AJ....123.2828C}, most closely match this comparison source, with the exception of a downturn in flux beyond 8100~{\AA} caused by poor flux calibration in the red and strong fringing. 
An M7 classification is also generally consistent with spectral index-based classifications using the VO1, VO2, TiO7 and Color-M indices defined in \citet[Table~\ref{tab:optindices}]{2003AJ....125.1598L}.  
Like {\namesh}, LP~704-48  exhibits no evidence of Li~I absorption (EW $<$ 0.13) or H$\alpha$ emission ($f_{H\alpha} < 3\times10^{-17}$~erg~s$^{-1}$~cm$^{-2}$; {\lha} $<$ $-$6.2)
in either our spectrum or that of \citet{2002AJ....123.2828C}. As discussed in Section~3, the lack
of nonthermal emission is consistent with a relatively old age for both sources.

\subsection{Low Resolution Near-Infrared Spectroscopy}

Low resolution near-infrared spectra of {\namesh} were obtained on 2009 November 4 (UT) with IRTF/SpeX in conditions of light cirrus and 0$\farcs$6 seeing. We used the prism-dispersed mode with the 0$\farcs$5 slit to obtain continuous, low-resolution ({\ldl} $\approx$ 120) spectra covering the 0.7--2.4~$\micron$ band.
Six exposures of 120~s each were acquired in an ABBA dither pattern along the slit which was aligned with the parallactic angle.
The A0 V star HD~1154 ($V$ = 8.85) was observed immediately afterward at a similar airmass for flux calibration and telluric absorption correction, and NeAr arc lamp and flat field exposures were obtained along with the standard.
Data were reduced with the SpeXtool package, version 3.4 \citep{2004PASP..116..362C,2003PASP..115..389V}, using standard settings.  A detailed description of our reduction procedures is given in \citet{2007AJ....134.1330B}.

\begin{figure}
\epsscale{0.8}
\plotone{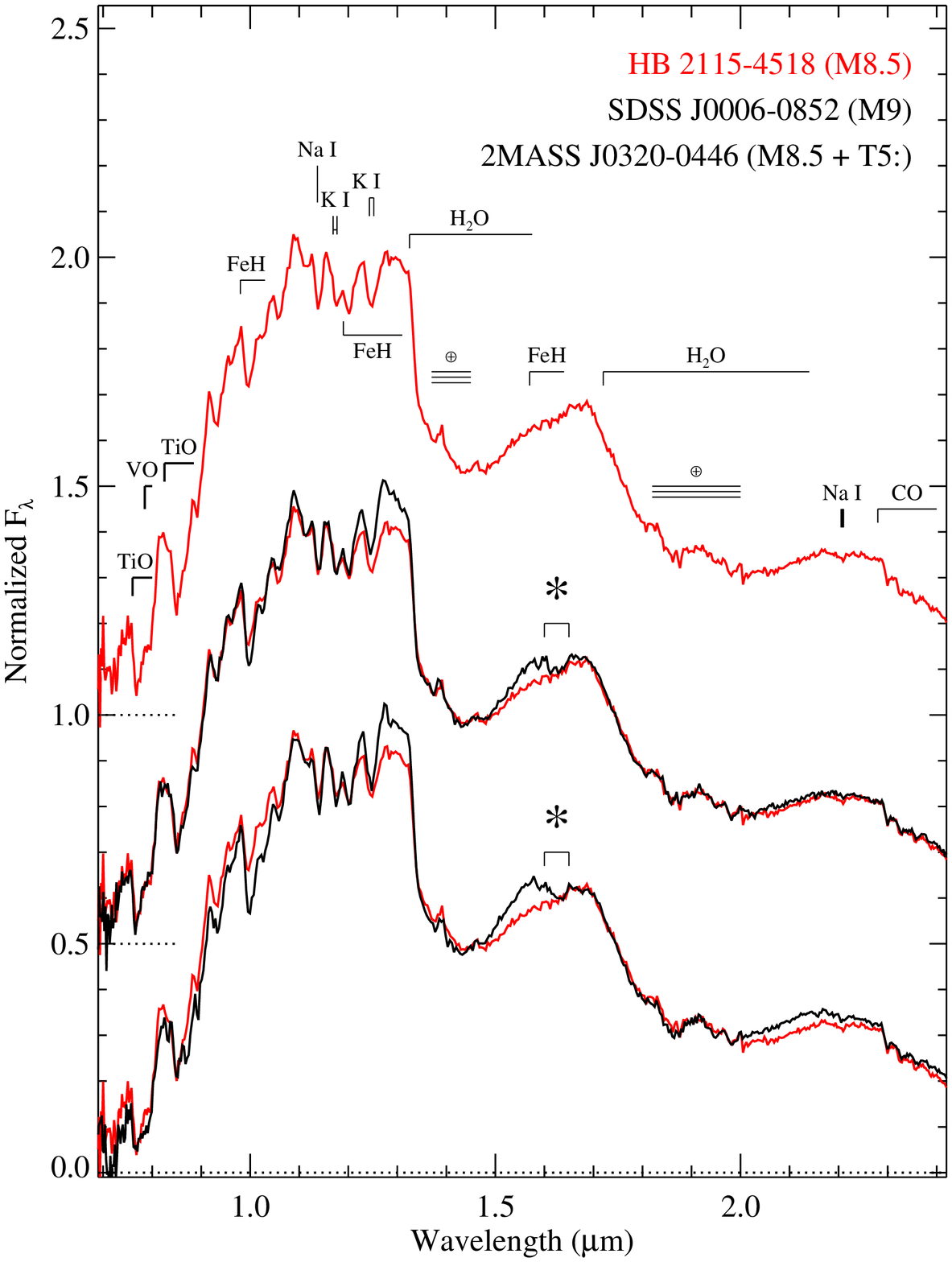} 
\caption{SpeX prism spectrum of {\namesh} (center) compared to those of the M8.5 HB~2115$-$4518 (top; \citealt{1988MNRAS.234..177H}; Burgasser et al., in prep.) and the M8.5+T5 spectral binary 2MASS~J0320$-$0446 (bottom; \citealt{2003IAUS..211..197W,2008ApJ...681..579B}).
All spectra are normalized in the 1.2--1.3~$\micron$ range and offset by constants.
The HB~2115$-$4518 spectrum (in red) is also overlaid on the other two spectra for comparison.  
Major molecular and atomic absorption features are labeled, as is the ``notch'' feature at 1.6~$\micron$ indicating the presence of an unresolved T dwarf companion.  Regions of strong telluric absorption are indicated by the horizontal lines.
\label{fig:spexspec}}
\end{figure}

Figure~\ref{fig:spexspec} displays the reduced spectrum of {\namesh} compared to equivalent data for 2MASS~J0320$-$0446 (SpeX data from \citealt{2008ApJ...681..579B}) and the M8.5 dwarf HB~2115$-$4518 (\citealt{1988MNRAS.234..177H,2005A&A...440.1061L}; SpeX data from Burgasser et al.~in prep.). 
The three spectra are generally equivalent in shape, exhibiting strong TiO, {\wat} and CO bands; weaker absorption from VO and FeH; and line absorption from K~I and Na~I (doublet lines are unresolved in these data).  The spectra of both {\namesh} and 2MASS~J0320$-$0446 also exhibit subtle features not present in that of HB~2115$-$4518, most notably an excess of flux at 1.27~$\micron$, 1.55~$\micron$ and (less prominently) 2.1~$\micron$; and a distinct ``notch'' feature at 1.6~$\micron$.  As discussed in \citet{2007AJ....134.1330B} and shown in Section~4, these features indicate the presence of an unresolved T dwarf companion to {\namesh}.

\subsection{High Resolution Near-Infrared Spectroscopy}

High resolution near-infrared spectra of {\namesh} and LP~704-48 were obtained on seven epochs between 2010 August and 2012 January using the NIRSPEC echelle spectrograph on the Keck II telescope.  Observations are summarized in Table~\ref{tab:nirspec}.  On four epochs we used the N3 order-sorting filter and 0$\farcs$432$\times$12$\arcsec$ slit to obtain 1.18--1.30~$\micron$ spectra over orders 56--66 with resolution {\ldl} = 20,000 ($\Delta{v}$ = 15~{\kms}) and dispersion of 0.181~{\AA}~pixel$^{-1}$. 
On 2011 September 7 and 2012 January 6 (UT) we observed {\namesh} and LP~704-48 with the N7 filter and 0$\farcs$432$\times$12$\arcsec$ slit to obtain 2.00--2.39~$\micron$ spectra over orders 32--38 with {\ldl} = 20,000 ($\Delta{v}$ = 15~{\kms}) and dispersion of 0.315~{\AA}~pixel$^{-1}$.  
On 2012 January 15 (UT) we observed {\namesh} with the N5 filter and 0$\farcs$288$\times$24$\arcsec$ slit to obtain 1.43--1.70~$\micron$ spectra over orders 45--53 with {\ldl} = 30,000 ($\Delta{v}$ = 10~{\kms}).
Echelle and cross-dispersion gratings were set to the values listed in Table~\ref{tab:nirspec}, which varied slightly from run to run to maintain consistent projection of arclamp images on the detector.
%These data were aimed at detecting reflex motion from the primary or secondary components or both, although in the end we deduce that only primary motion could be detected.  
We also observed seven L dwarfs with previously measured radial velocities from \citet{2010ApJ...723..684B} to serve as radial velocity standards (Table~\ref{tab:nirspec_cals}).  For each source, we obtained spectra in AB or ABBA nodding sequences.  We also observed a nearby A0~V star for flux calibration and telluric correction, with the exception of some N5 observations for which a comparison star was not required (see below).  Dark, quartz lamp and NeArXeKr arc lamp frames were obtained at the beginning or end of each night without changing the instrument configuration.

\begin{deluxetable*}{llcccccccll}
\tablecaption{NIRSPEC Observations of {\namesh} and LP~704-48 \label{tab:nirspec}}
\tabletypesize{\scriptsize}
%\rotate
\tablewidth{0pt}
\tablehead{
\colhead{Source} &
\colhead{Date (UT)} &
\colhead{MJD} &
\colhead{Filter} &
\colhead{$\lambda$ ($\micron$)} &
\colhead{Slit} &
\colhead{Echelle/Grating ($\degr$)} &
\colhead{T (s)} &
\colhead{Airmass} &
\colhead{Calibrator} &
\colhead{Conditions} \\
}
\startdata
{\namesh} & 2010 Aug 19 & 55437.5 & N3 & 1.18--1.30 & 0$\farcs$432$\times$12 & 62.95/34.10 & 3600 & 1.13--1.20 & HD~3604 & clear, seeing $\approx$ 0$\farcs$6  \\
{\namesh} & 2010 Nov 26 & 55526.3 & N3 & 1.18--1.30 & 0$\farcs$432$\times$12 & 62.95/34.10 & 4000 & 1.14--1.33 & HD~3604 & high humidity, seeing $\approx$ 1$\farcs$5  \\
{\namesh} & 2011 Jul 06 & 55748.6 & N3 & 1.18--1.30 & 0$\farcs$432$\times$12 & 62.95/34.08 & 1500 & 1.14--1.16 & HD~3604 & clear, seeing $\approx$ 0$\farcs$7  \\
{\namesh} & 2011 Sep 07 & 55811.5 & N7 & 2.00--2.39 & 0$\farcs$432$\times$12 & 63.00/35.46 & 2000 & 1.25--1.37 & HD~13936 & light cirrus, seeing $\approx$ 1$\arcsec$ \\
{\namesh} & 2011 Sep 10 & 55814.5 & N3 & 1.18--1.30 & 0$\farcs$432$\times$12 & 62.95/34.08 & 2000 & 1.24--1.41 & HD~1154 & clear, seeing $\approx$ 1$\farcs$5  \\
{\namesh} & 2012 Jan 6 & 55932.2 & N7 & 2.00--2.39 &  0$\farcs$432$\times$12 & 63.01/35.47 & 2000 & 1.24--1.35 & HD~1154 & clear, seeing $\approx$ 0$\farcs$5  \\
LP 704$-$48 & 2012 Jan 6 & 55932.2 & N7 & 2.00--2.39 &  0$\farcs$432$\times$12 & 63.01/35.47 & 800 & 1.19--1.20 & HD~1154 & clear, seeing $\approx$ 0$\farcs$5  \\
{\namesh} & 2012 Jan 15 & 55941.2 & N5 & 1.43--1.70 & 0$\farcs$288$\times$24 & 63.04/36.30 & 2400 & 1.28--1.40 & 48 Cet & light clouds, seeing $\approx$ 0$\farcs$8  \\
\enddata
\end{deluxetable*}

\begin{deluxetable*}{lllcccccc}
\tablecaption{NIRSPEC Observations of Radial Velocity Calibrators}
\label{tab:nirspec_cals}
\tabletypesize{\scriptsize}
\tablewidth{0pt}
\tablehead{
\colhead{Source} &
\colhead{Optical} &
\colhead{Obs.\ Date} &
\colhead{Filter} &
\colhead{Integration} &
\colhead{Airmass} &
\colhead{Calibrator} &
\colhead{RV\tablenotemark{a}} &
\colhead{Refs} \\
 & 
\colhead{SpT} & 
\colhead{(UT)} & &
\colhead{(s)} & & & 
\colhead{({\kms})} &
 \\
}
\startdata
%2MASS~J00361617+1821104 & L3.5 & 2011 Sep 10 & N7 & INT & AM & CAL & 19.02$\pm$0.15 & BLA10 \\ 
% &  &  XXXXX & N5 & XXX & XXX & \nodata\tablenotemark{b} \\
%BRI~0021-0214 & M9.5 & XXXXX & N5 & XXXX & XXXX & XXXXX & 4.4$\pm$2.2 & BAS06 \\
2MASS~J05233822$-$1403022 & L2.5 & 2009 Oct 31 & N3 & 2000 & 1.20--1.22 & $\tau$ Lep & 12.21$\pm$0.09 &  1 \\
%2MASS~J07003664+3157266 & L3.5 & 2011 Mar 18 & N3 & INT & AM & CAL & -42.42$\pm$0.09 &  BLA10 \\
2MASS~J09211410$-$2104446 & L1.5 & 2011 Oct 26 & N3 & 1200 & 1.32--1.33 & HD 82724 & 80.53$\pm$0.11 & 2 \\ 
2MASS~J10220489+0200477 & L1 & 2011 Mar 18 & N7 & 1800 & 1.28 & 39 Uma & 19.29$\pm$0.11 &  2 \\
%2MASS 1507 & L5 & 2011 Jun 10 & N7 &  INT & AM & CAL & -39.85$\pm$0.05 &  BLA10 \\
2MASS~J15150083+4847416 & L6 & 2011 Sep 10 & N7 & 350 & 1.57--1.63 & HD 143187 & $-$29.97$\pm$0.14 &  3 \\
DENIS-P~J170548.38$-$051645.7 & L4 & 2011 Aug 11 & N7 & 1200 & 1.24 & HD 159415 & 12.19$\pm$0.11 &  4 \\
2MASS~J17312974+2721233 & L0 & 2011 Jun 10 & N7 & 1000 & 1.03 & HD 165029 & $-$29.76$\pm$0.11 &  2 \\
2MASS~J22244381$-$0158521 & L4.5 & 2011 Sep 07 & N7 & 1500 & 1.08-1.11 & HD 198070 & $-$37.55$\pm$0.09 &  5 \\
%LHS 2065 & M9 & XXXX & N5 & XXXX & XXXX & XXXX & 8.7$\pm$.15 & TIN98; IN PRATO02 \\
\enddata
\tablenotetext{a}{Radial velocity measurements from \citet{2010ApJ...723..684B}.}
\tablerefs{
(1) \citet{2003AJ....126.2421C}; 
(2) \citet{2008AJ....136.1290R}; 
(3) \citet{2003IAUS..211..197W}; 
(4) \citet{2004A&A...416L..17K}; 
(5) \citet{2000AJ....120..447K}.}
\end{deluxetable*}

All data were reduced using REDSPEC, an IDL-based software package developed at UCLA for NIRSPEC by S.\ Kim, L.\ Prato, and I.\ McLean\footnote{See \url{http://www2.keck.hawaii.edu/inst/nirspec/redspec/index.html}.}.  Images were first corrected for pixel-response variations using the dark-subtracted flat field frames.  Individual orders were then isolated, rectified, and pair-wise subtracted, and spectra were extracted by summing across rows.  In this study, we focused exclusively on order 59 (1.283--1.300~$\micron$) in the N3 data, order 49 (1.545--1.567~$\micron$) in the N5 data, and order 33 (2.291--2.326~$\micron$) in the N7 data.  The latter two orders are relatively devoid of telluric absorption features \citep{2002ApJ...569..863P,2007ApJ...658.1217M}, while order 33 samples the strong CO band around 2.3~$\micron$ \citep{2008ApJ...678L.125B}.  Image rectification and wavelength calibration were performed using telluric OH emission lines present in the long science exposures for orders 49 and 59 \citep{2000A&A...354.1134R}; for order 33, we used the arc lamp images and a vacuum line list from the National Institute of Standards and Technology (NIST) atomic spectral line database \citep{NIST}.  For each observation, wavelength solutions were corrected for barycentric motion.  Flux calibration and telluric absorption corrections (except for order 49) were calculated from the A0~V spectra assuming a 9480~K blackbody.  No filtering of fringing was performed.  

\begin{figure}[h]
\epsscale{1.0}
\plotone{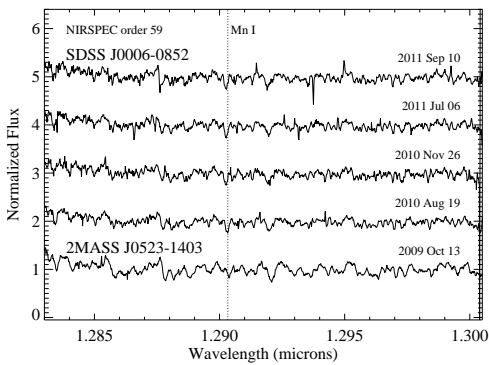} 
\caption{Four epochs of NIRSPEC order 59 (N3) spectra of {\namesh} (top) compared to the L2.5 radial velocity standard 2MASS~J0523$-$1403 (bottom). 
Spectra have been corrected for barycentric motion, normalized to the median flux and offset by integer constants. The 1.290~$\micron$ Mn~I line is labelled; other features arise primarily from FeH and {\wat} absorption.
%[QUESTION: SHOULD I LABEL FEH AND H2O TRANSITIONS?]
\label{fig:nirspec1}}
\end{figure}

\begin{figure}[h]
\plotone{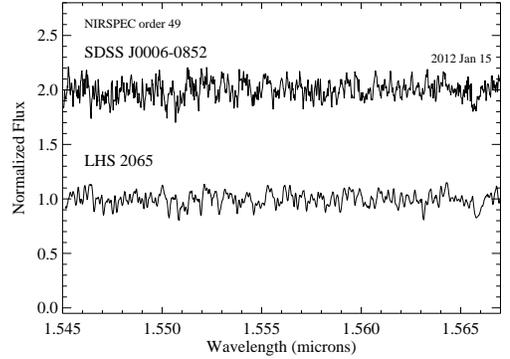} 
\caption{NIRSPEC order 49 (N5) spectrum of {\namesh} obtained on 2012 Jan 15 (UT, top) compared to that of the M9 radial velocity standard LHS~2065 (bottom; from \citealt{2002ApJ...569..863P}).  
Spectra have been corrected for barycentric motion, normalized to the median flux and offset by integer constants. 
Features arise primarily from {\wat} and FeH absorption.
%[QUESTION: SHOULD I LABEL FEH AND H2O TRANSITIONS?]
\label{fig:nirspec2}}
\end{figure}

\begin{figure}[h]
\plotone{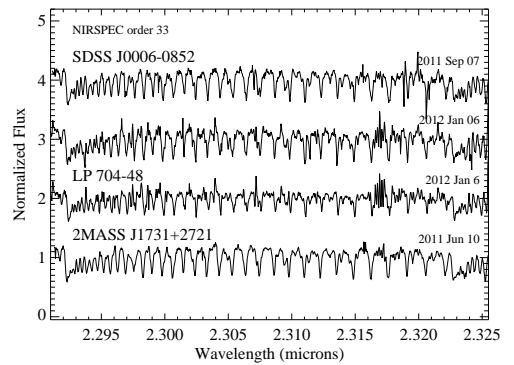} 
\caption{NIRSPEC order 33 (N7) spectra of {\namesh} (top two spectra), LP~704-48 (second from the bottom) and the L0 radial velocity standard 2MASS~J1731+2721 (bottom).  
Spectra have been corrected for barycentric motion, normalized to the median flux and offset by integer constants. 
Features arise primarily from CO absorption.
%[QUESTION: SHOULD I LABEL CO TRANSITIONS?]
\label{fig:nirspec3}}
\end{figure}

Figures~\ref{fig:nirspec1} through~\ref{fig:nirspec3} display all of the reduced spectral data for {\namesh} and LP~704-48 compared to select radial velocity standards.
Signal-to-noise (S/N) for these data are generally $>$50, with the exception of our N5 data which has S/N $\approx$ 20.
All of the spectra show numerous molecular transitions arising from {\wat} and FeH in orders 59 and 49, and CO in order 33.  We also detect the weak Mn~I line at 1.2903~$\micron$ in the spectrum of {\namesh} \citep{2007ApJ...658.1217M}.

\section{LP~704-48 and {\namesh}}

\subsection{Evidence for Common Proper Motion and Distance}

Our Nickel astrometry (Figure~\ref{fig:nickelpm}) shows convincingly that LP~704-48 and {\namesh} comprise a common proper motion pair.  To quantify this, we combined our measurements with prior astrometry from the SuperCosmos Sky Survey \citep{2001MNRAS.326.1279H,2001MNRAS.326.1295H,2001MNRAS.326.1315H}, the 2MASS All-Sky Point Source Catalog, and SDSS Data Release 7 \citep{2009ApJS..182..543A}. Linear fits yield proper motions of [$\mu_{\alpha}\cos{\delta}$,$\mu_{\delta}$] = 
[$-$70$\pm$3,$-$314$\pm$2]~{\masyr} and 
[$-$84$\pm$28,$-$337$\pm$17]~{\masyr} for LP~704-48 and {\namesh}, respectively.
The former agrees with values reported by \citet{2003ApJ...582.1011S}.
These motions are consistent with each other to within 0.5$\sigma$ in Right Ascension and 1.3$\sigma$ in declination.  The combined astrometry also yield angular separations of 25$\farcs$26$\pm$0$\farcs$09 in Right Ascension and $-$10$\farcs$64$\pm$0$\farcs$17 in declination, pointing from LP~704-48 to {\namesh}.

\begin{deluxetable}{lccc}
\tabletypesize{\scriptsize}
%\rotate
\tablecaption{Distance Estimates for {\namesh} and LP~704-48 \label{tab:distance}}
\tablewidth{0pt}
\tablehead{
 & \multicolumn{3}{c}{Distance (pc)} \\
 \cline{2-4}
\colhead{Method} &
\colhead{LP~704-48} &
\colhead{\namesh} &
\colhead{$\Delta/\sigma$} \\
}
\startdata
H02 $M_i$ vs $i-z$ & 28$\pm$5 & 31$\pm$6 & -0.4 \\
H02 $M_i$ vs $i-J$ & 31$\pm$6 & 30$\pm$6 & +0.1 \\
H02 $M_J$ vs SpT\tablenotemark{a} & 17$\pm$4 & 36$\pm$7 & -2.4 \\
C03 $M_J$ vs SpT\tablenotemark{a} & 18$\pm$3 & 35$\pm$4 & -3.4 \\
W05 $M_i$ vs $i-z$ & 17$\pm$2 & 33$\pm$5 & -3.0 \\
W05 $M_i$ vs $i-J$ & 27$\pm$3 & 31$\pm$3 & -0.3 \\
B10 $M_r$ vs $r-z$ & 22$\pm$4 & \nodata\tablenotemark{b} & \nodata \\
B10 $M_r$ vs $r-i$ & 21$\pm$4 & \nodata\tablenotemark{b} & \nodata  \\
B10 $M_r$ vs $i-z$ & 26$\pm$6 & \nodata\tablenotemark{b} & \nodata  \\
S10 $M_i$ vs $i-z$ & \nodata\tablenotemark{b} & 31$\pm$8 & \nodata  \\
S10 $M_i$ vs $i-J$ & \nodata\tablenotemark{b} & 36$\pm$6  & \nodata \\
D12 $M_J$ vs SpT\tablenotemark{a} & 20$\pm$5 & 37$\pm$8 & -1.8 \\
D12 $M_H$ vs SpT\tablenotemark{a} & 20$\pm$4 & 40$\pm$8 & -2.2  \\
D12 $M_{K_s}$ vs SpT\tablenotemark{a} & 20$\pm$4 & 42$\pm$9 & -2.2  \\
\cline{1-4} 
Weighted Mean & 22$\pm$5 & 32$\pm$2 & -2.0 \\
\enddata
\tablenotetext{a}{Excluded from final average.}
\tablenotetext{b}{Outside defined color range.}
\tablerefs{(H02): \citet{2002AJ....123.3409H} ; (C03): \citet{2003AJ....126.2421C} ; (W05): \citet{2005PASP..117..706W}; (B10): \citet{2010AJ....139.2679B}; (S10): \citet{2010AJ....139.1808S}; (D12): \citealt{2012ApJS..201...19D}}
\end{deluxetable}

We also find that the estimated distances to these sources are in formal agreement, based on a combination of several optical and near-infrared absolute magnitude/color and absolute magnitude/spectral type relationships\footnote{Note that linear coefficient of $M_r$ versus $r-i$ in Table~4 of \citet{2010AJ....139.2679B} should be +4.548, not $-$4.548.} defined in
\citet{2002AJ....123.3409H,2003AJ....126.2421C,2005PASP..117..706W,2010AJ....139.1808S,2010AJ....139.2679B}; and \citet{2012ApJS..201...19D}.
Table~\ref{tab:distance} summarizes these estimates, which incorporate uncertainties in photometry, spectral classifications (0.5~subtypes), and the reported systematic uncertainty in the absolute magnitude relations.
Some relationships show better agreement between the two components than others, with the absolute magnitude/spectral type relations diverging the most.  Using only the color relations, we find mean distances of 22$\pm$5~pc for LP~704-48 and 32$\pm$2~pc for {\namesh}, a 2$\sigma$ difference.  Note that nearly all of the relations place LP~704-48 roughly 50\% closer, suggesting that it to may be an unresolved binary, although age/surface gravity effects cannot be ruled out (see below).  We adopt a weighted mean of 30$\pm$3~pc for the combined system, implying a projected separation of 820$\pm$120~AU.

\subsection{Probability of Association}

To assess the probability of chance alignment  for this wide pairing, we followed the method described in \citet{2010AJ....139.2566D}, which estimates the frequency of unrelated pairings using a Galactic model based on an empirical luminosity function \citep{2008ApJ...673..864J,2010AJ....139.2679B} and an empirical space velocity distribution \citep{2007AJ....134.2418B}.
The number of single (and hence unrelated) stars within a 6D ellipsoid defined by the angular separation of the binary, the estimated distance to the binary, the space motions of the binary, and the uncertainties in these values was determined through Monte Carlo simulation.   From 10$^6$ simulations, we found that only 0.0015 stars were spatially coincident and had proper motions similar to the values observed for this wide pair. When the (far more precise) radial motions of {\namesh} and LP~704-48 were also considered (see below), the number of chance alignments fell to zero (i.e. probability $<$10$^{-6}$). We therefore conclude with high confidence that LP~704-48 and {\namesh} are not a chance alignment. 

\subsection{Activity, Age and Metallicity}

Additional constraints on the physical properties of this system can be inferred from our optical spectroscopy of LP~704-48 and {\namesh}.
The lack of Li~I absorption in either source rules out masses below 0.06~{\msun}, implying ages older than 90~Myr and 210~Myr, respectively, based on the evolutionary models of \citet{2003A&A...402..701B} and {\teff} estimates of 2660$\pm$150~K and 2400$\pm$160~K as inferred from the {\teff}/SpT relation of
\citet{2009ApJ...702..154S}.\footnote{Uncertainties include classification of 0.5 subtypes for both components and a 100~K systematic uncertainty in the \citet{2009ApJ...702..154S} relation.}  
More stringent age constraints come from the lack of H$\alpha$ emission in either source.
For {\namesh}, one could attribute the lack of emission to an increasingly neutral photosphere that is decoupled from the magnetic field, a hypothesis used to explain the decline in both H$\alpha$ and X-ray emission in dwarfs later than M8--M9 (e.g., \citealt{2000AJ....120.1085G,2002ApJ...571..469M,2004AJ....128..426W}).  However, the inactivity of LP~704-48 is remarkable given the high incidence of H$\alpha$ emission
among M7 dwarfs in the vicinity of the Sun ($\gtrsim$90\% of sources with vertical scaleheights $|Z| <$ 50~pc; \citealt{1996AJ....112.2799H,2000AJ....120.1085G,2004AJ....128..426W,2006AJ....132.2507W,2008AJ....135..785W}). From the activity frequencies of M dwarfs at various scaleheights, \citet{2008AJ....135..785W} have inferred an activity lifetime for M7 dwarfs of 8$^{+0.5}_{-1.0}$~Gyr. 
Since emission from LP~704-48 is seen in neither our spectrum nor that of \citet{2002AJ....123.2828C}, and given that the limit on H$\alpha$ luminosity is nearly two orders of magnitude below the local active mean ({\lha} = $-$4.3;  \citealt{2004AJ....128..426W}), we conclude that this source is truly inactive and that the LP~704-48/{\namesh} system is likely to be quite old.  

\begin{deluxetable*}{lcccl}
\tablecaption{Properties of the LP 704-48/{\namesh}AB System \label{tab:properties}}
\tabletypesize{\scriptsize}
\tablewidth{0pt}
\tablehead{
\colhead{Parameter} &
\colhead{LP~704-48} &
\colhead{{\namesh}AB} &
\colhead{System} &
\colhead{Ref.} \\
}
\startdata
Optical SpT & M7 & M9  & \nodata & 1,2,3 \\
NIR SpT & \nodata &  M8.5$\pm$0.5 + T5$\pm$1  &  \nodata & 1 \\
$r$ &  17.313$\pm$0.005 & {21.04$\pm$0.06} & \nodata & 4 \\
$r-i$ &  2.373$\pm$0.006 & {2.82$\pm$0.06} & \nodata & 4 \\
$i-z$ &  1.197$\pm$0.006 & {1.77$\pm$0.03} &  \nodata & 4 \\
$i-J$ &  1.78$\pm$0.02 & {2.32$\pm$0.04} &  \nodata & 4,5 \\
$J$ &  11.97$\pm$0.02 & {14.14$\pm$0.04} & \nodata & 5 \\
$J-K_s$ &  0.88$\pm$0.03 & {1.01$\pm$0.05} & \nodata & 5 \\
{\lha} & $< -6.2$ & $< -5.7$ & \nodata & 1 \\
Est.\ $d$ (pc) & 22$\pm$5 & 32$\pm$2 & 30$\pm$3 &  1 \\
$\mu_{\alpha}$ (mas~yr$^{-1}$) & $-$70$\pm$3 & $-$84$\pm$28 & $-$70$\pm$3 & 1 \\
$\mu_{\delta}$ (mas~yr$^{-1}$) & $-$314$\pm$2 & $-$337$\pm$17 &  $-$315$\pm$2 & 1 \\
RV ({\kms}) & $-$15.3$\pm$0.9 & Variable & $-$15.3$\pm$0.3\tablenotemark{a} & 1 \\
$a$ (AU) & \nodata & 0.286$\pm$0.009 & 820$\pm$120 & 1 \\
$a$ ($\arcsec$) & \nodata & 0$\farcs$0095$\pm$0$\farcs$0010\tablenotemark{b} & 27$\farcs$41$\pm$0$\farcs$10 & 1 \\
$U$ ({\kms}) & \nodata & \nodata &  31$\pm$3 & 1 \\
$V$ ({\kms}) & \nodata & \nodata &  $-$37$\pm$3 & 1 \\
$W$ ({\kms}) & \nodata & \nodata & 2$\pm$2 & 1 \\
Est.\ Age (Gyr) & $\gtrsim$8 & $\gtrsim$3--4 & $\gtrsim$3--4 & 1,2 \\
Est.\ Masses ({\msun})\tablenotemark{c} & 0.092 & 0.082--0.083, 0.049--0.064 & 0.22--0.24 & 1 \\
\enddata
\tablenotetext{a}{Based on constrained radial velocity orbit fit to {\namesh}AB; see Table~\ref{tab:orbit}.}
\tablenotetext{b}{At maximum elongation.}
\tablenotetext{c}{For an age range of 3--10~Gyr, based on the evolutionary models of  Baraffe et al.~(2003).}
\tablerefs{
(1) This paper;
(2) \citet{2008AJ....135..785W}; 
(3) \citet{2002AJ....123.2828C}; 
(4) SDSS: \citet{2000AJ....120.1579Y}; 
(5) 2MASS: \citet{2006AJ....131.1163S}.}
\end{deluxetable*}

This conclusion is supported by the kinematics of the system.  Combining our proper motion measurement, distance estimate and systemic radial velocity determination below, we derive the heliocentric $UVW$ velocities listed in Table~\ref{tab:properties}.  These velocities are on the boundary of the young-old disk as defined in \citet{1992ApJS...82..351L}.  
The disk classification is consistent with what appear to be near-solar metallcities for LP~704-48 and {\namesh}, as evident from their spectral energy distributions and as quantified by the $\zeta$ index of \citet{2007ApJ...669.1235L}.  Both sources have $\zeta \approx 1.00$ (Table~\ref{tab:optindices}) indicating roughly solar metallicities.

We therefore conclude that LP~704-48 and {\namesh} comprise a physically associated, widely-separated system of relatively old, VLM stars with near-solar metallicity. 

\section{The M Dwarf/T Dwarf Spectral Binary {\namesh}}

As discussed in Section~2.1, the peculiar features observed in the low-resolution near-infrared spectrum of {\namesh} indicate the presence of an unseen T dwarf companion.
To test this hypothesis, we performed a spectral template fitting analysis similar to that described in \citet{2010ApJ...710.1142B}.
We drew 638 spectra of 614 sources with S/N $\gtrsim$ 20  from the SpeX Prism Spectral Libraries\footnote{See \url{http://www.browndwarfs.org/spexprism}. Data were drawn from \citet{2004AJ....127.2856B,2006ApJ...639.1095B, 2006ApJ...637.1067B,2007ApJ...658..557B,2008ApJ...681..579B,2008ApJ...674..451B,2010ApJ...710.1142B,2004ApJ...604L..61C,2006AJ....131.1007B,2006AJ....131.2722C,2006AJ....132.2074M,2006ApJ...639.1114R,2007ApJ...659..655B,2007AJ....134.1330B,2007ApJ...658..617B,2007ApJ...655..522L,2007AJ....134.1162L,2008ApJ...686..528L,2007ApJ...654..570L,2007AJ....133.2320S,2009AJ....137..304S,2010ApJS..190..100K}; Cruz et al.~in prep.; and Burgasser et al.~in prep.} with published optical and/or near-infrared spectral types between M7 and T8, excluding known binaries, subdwarfs and low-surface gravity brown dwarfs.  The spectra were reclassified in the near-infrared using the index-based scheme defined in \citet{2007ApJ...659..655B}.
Spectral fluxes were then scaled to absolute $F_{\lambda}$ units using the $M_J$/spectral type relation of \citet{2003AJ....126.2421C} for types M7--L2 and the $M_{K_s}$/spectral type relation of \citet{2008ApJ...685.1183L} for types L2--T8.  We then combined pairs of flux-calibrated spectra to create binary spectral templates ($B(\lambda)$), constraining the spectral types of the primary to M7--L8 and of the secondary to L9--T8, thereby generating 44605 unique binary templates.  We compared both the original 638 single spectra and the binary templates to the spectrum of {\namesh} ($S({\lambda})$) over the wavelength ranges $\{\lambda\}$ = 0.95--1.35~$\micron$, 1.45--1.80~$\micron$ and 2.00--2.35~$\micron$ using the $\chi^2$ statistic,
\begin{equation}
\chi^2 = \sum_{\{\lambda\}} \left(\frac{B(\lambda)-S(\lambda)}{\sigma(\lambda)}\right)^2
\end{equation}
where $\sigma(\lambda)$ corresponds to the uncertainty spectrum of {\namesh} alone.

\begin{figure*}[h]
\epsscale{1}
\plottwo{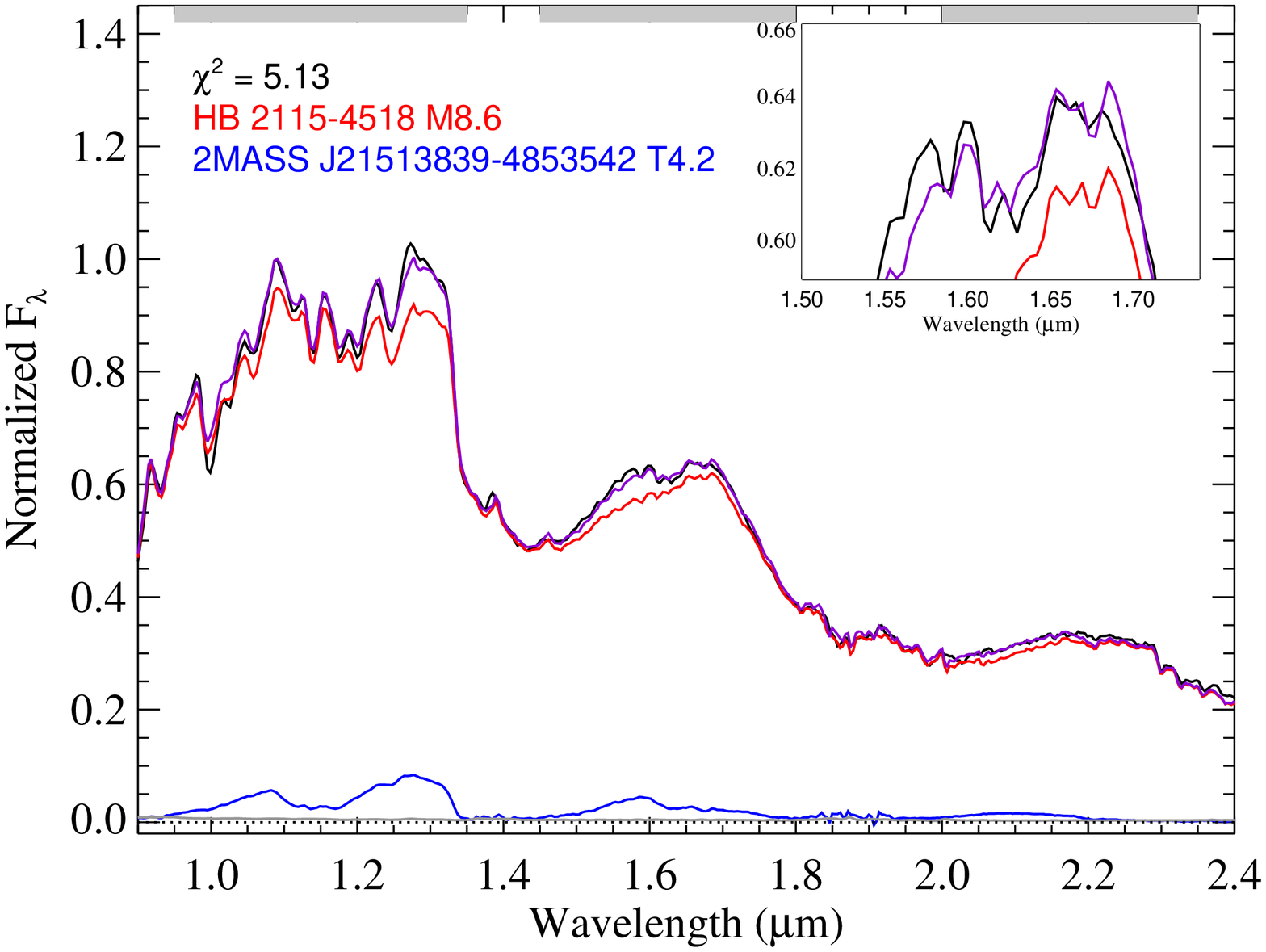}{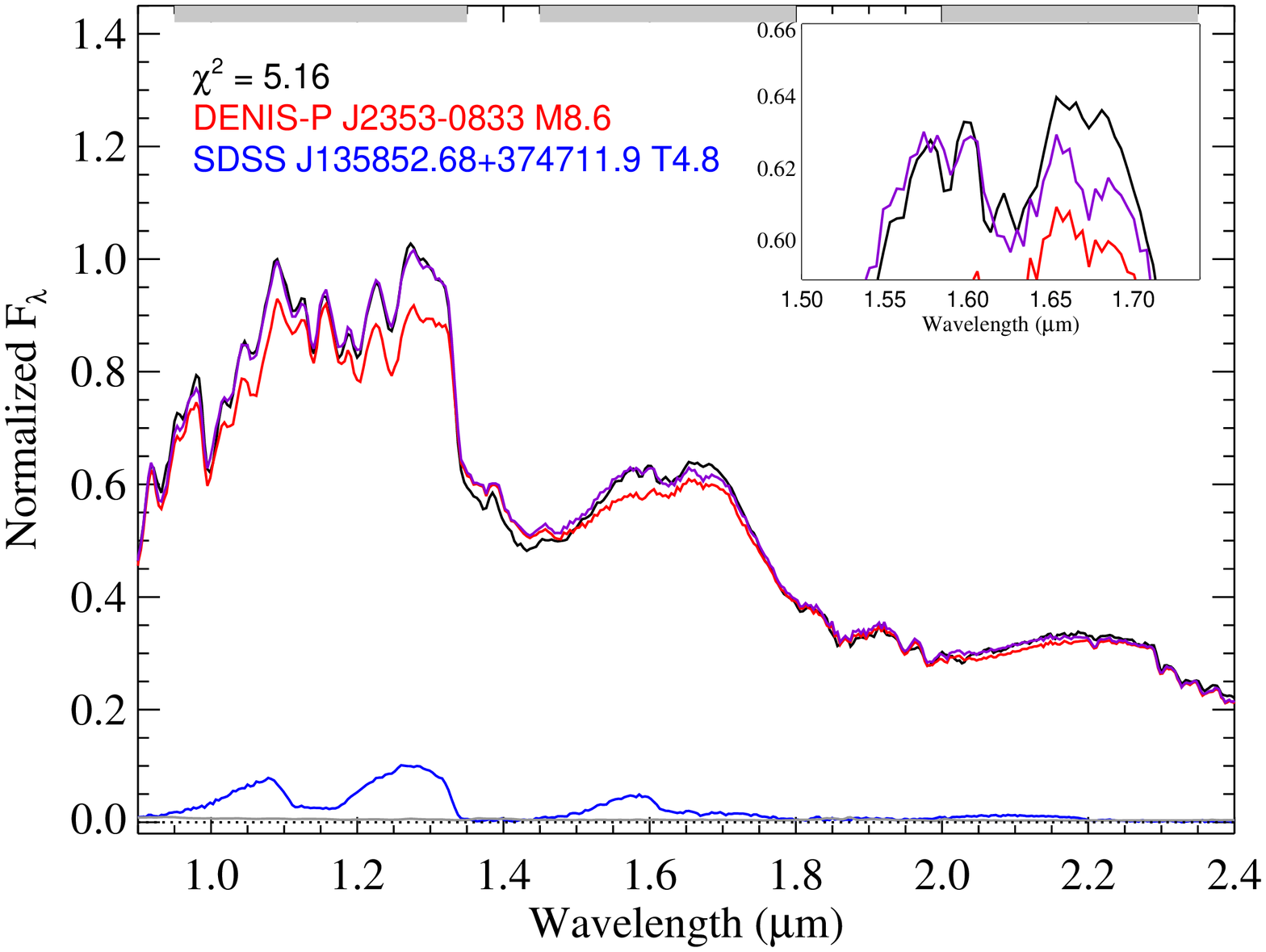} \\
\plottwo{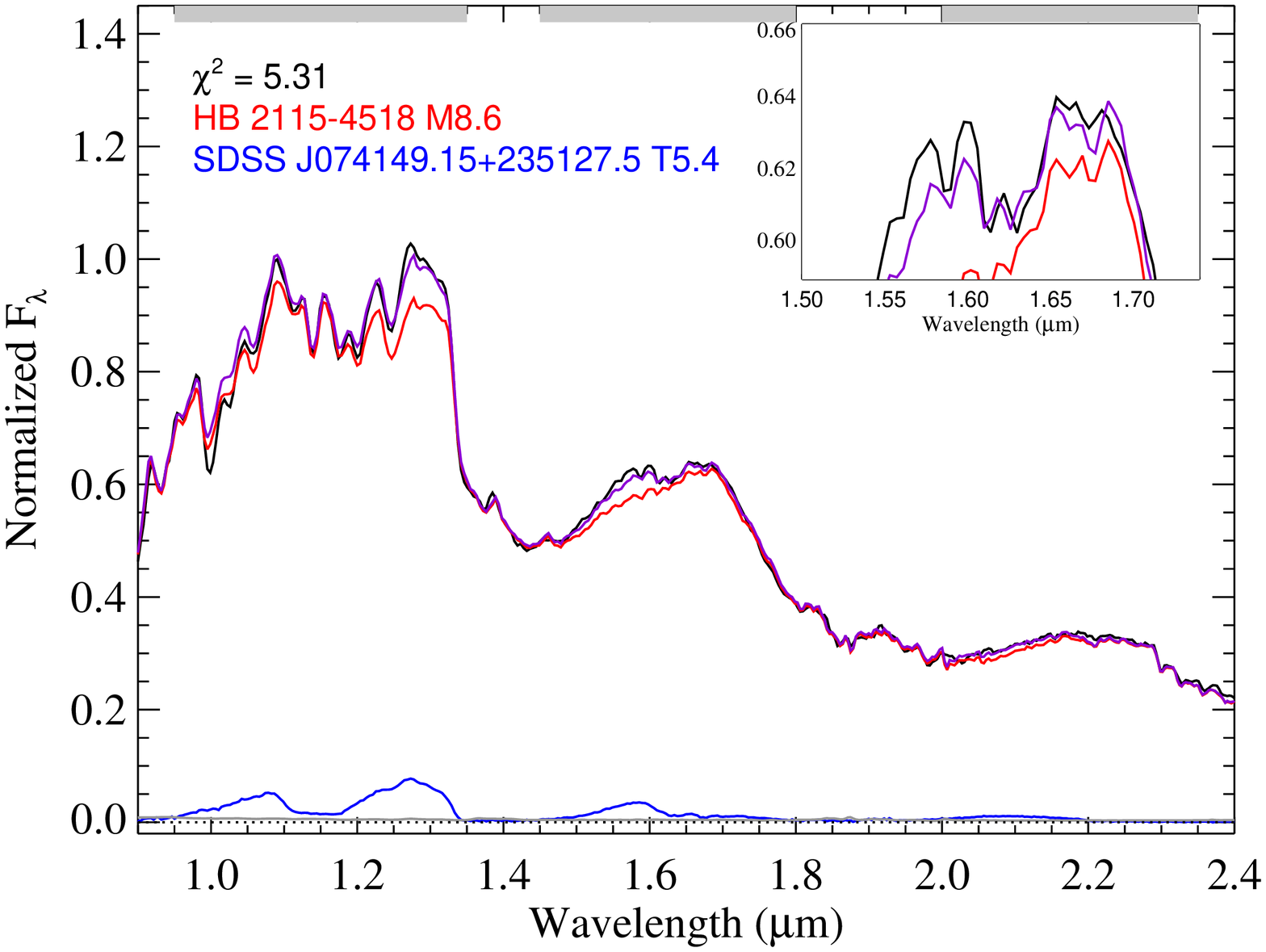}{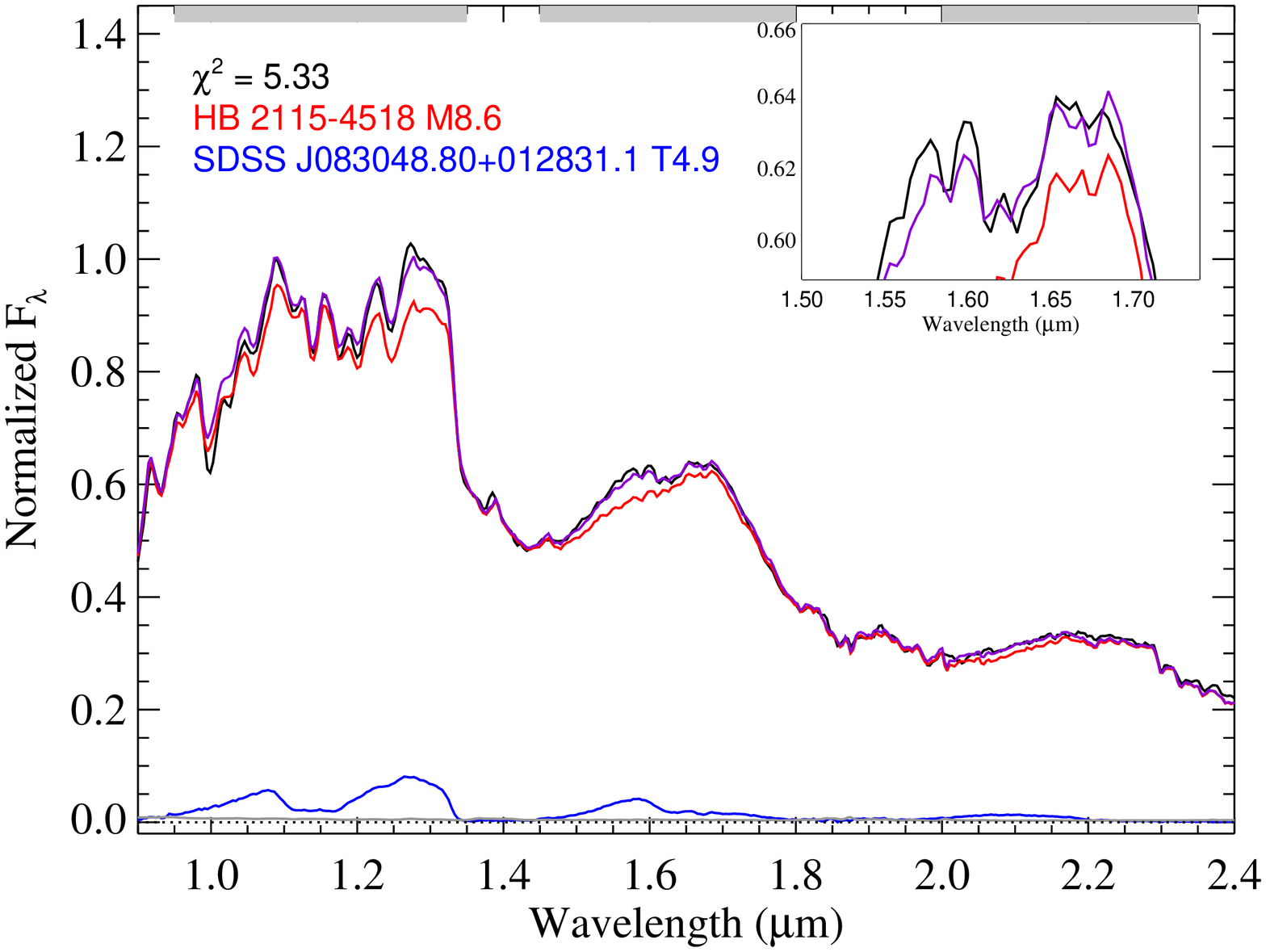} \\
\caption{SpeX spectrum of {\namesh} (black lines) compared to the four best-fitting binary spectral templates (purple lines), all normalized in the 1.1--1.3~$\micron$ region.  Primary (red lines) and secondary (blue lines) component spectra for the spectral binaries are also shown, scaled according to their contribution to the combined-light spectrum (source names in each panel are listed). The uncertainty spectrum of {\namesh} is indicated as the grey line at bottom.  Shaded regions at the top of the panels indicate the spectral ranges over which the spectrum of {\namesh} and the binary templates were compared, and $\chi^2$ values are listed.   The inset boxes show a close up of the 1.50--1.75~$\micron$ region where the notch feature is detected.
\label{fig:spexfit}}
\end{figure*}

\begin{deluxetable*}{lclccccc}
\tabletypesize{\scriptsize}
%\rotate
\tablecaption{Best-Fitting Spectral Binary Templates for {\namesh}. \label{tab:spexbinaryfit}}
\tablewidth{0pt}
\tablehead{
\colhead{Primary } & \colhead{SpT} & \colhead{Secondary } & \colhead{SpT} & \colhead{ ${\Delta}J_{MKO}$} & \colhead{ ${\Delta}H_{MKO}$} & \colhead{ ${\Delta}K_{MKO}$} & \colhead{$\chi^2$}}
\startdata
HB 2115-4518	& M8.6 & 2MASS J21513839-4853542 &	T4.2 & 2.886 &	3.384 &	3.908 & 5.13\\
DENIS-P J2353-0833 &	M8.6 & SDSS J135852.68+374711.9 &	T4.8 & 2.930 &	2.667 &	3.471 & 5.16 \\
HB 2115-4518	& M8.6 & SDSS J074149.15+235127.5 &	T5.4 & 3.096 &	3.880 &	4.457 & 5.31 \\
HB 2115-4518 & M8.6 & SDSS J083048.80+012831.1 & T4.9 & 2.953 &	3.598 &	4.168 & 5.33\\
HB 2115-4518	& M8.6 & 2MASSI J2254188+312349 &	T3.9 & 3.097 &	3.455 &	3.923 & 5.50 \\
DENIS-P J2353-0833 &	M8.6 & 2MASSI J0937347+293142 & T5.6 & 3.199	& 3.086 & 	3.981 & 5.53 \\
DENIS-P J2353-0833 &	M8.6 & SDSS J083048.80+012831.1 &	T4.9 & 3.285 &	2.950 &	3.607 & 5.54 \\
HB 2115-4518 & M8.6 & SDSS J074201.41+205520.5 & T5.2 & 3.107 &	3.838 &	4.492 & 5.54 \\
HB 2115-4518 & M8.6 & 2MASS J05591914-1404488 & T4.4 & 3.019 &	3.615 &	4.176 & 5.56 \\
HB 2115-4518 & M8.6 & 2MASS J06020638+4043588 & T4.6 & 2.986 &	 3.637 &	4.200 & 5.57 \\
\tableline
{\bf Average Primary} & M8.7$\pm$0.2 & {\bf Average Secondary} & T4.8$\pm$1.1 & 3.2$\pm$0.3 & 3.8$\pm$0.5 & 4.4$\pm$0.6 & 99.8\%\tablenotemark{a} 
\enddata
\tablenotetext{a}{Confidence level that the best-fit binary template provides a statistically better match than the best-fit single template (HB 2115-4518, $\chi^2 = 12.6$) based on the F-test statistic.}
\end{deluxetable*}

Figure~\ref{fig:spexfit} displays the four best-fitting binary templates (the best-fitting single template, HB~2115$-$4518, is shown in Figure~\ref{fig:spexspec}).  A combination of a late-type M dwarf and mid-type T dwarf accurately reproduces the spectrum of {\namesh}, including the excess flux at 1.27~$\micron$, 1.55~$\micron$ and 2.1~$\micron$.  This combination also reproduces with reasonable fidelity the notch feature at 1.62~$\micron$,
arising from FeH absorption in the M dwarf primary and the pseudo-continuum peak blueward of the 1.6~$\micron$ {\meth} band in the T dwarf secondary \citep{2007AJ....134.1330B}.  Several dozen binary templates were found to provide statistically superior matches to the spectrum of {\namesh} as compared to HB~2115$-$4518 alone ($\chi^2$ = 12.5), with significance values $>$99\% based on the F-test statistic (see Eqns.~2--6 in \citealt{2010ApJ...710.1142B}).  
Table~\ref{tab:spexbinaryfit} details the ten best binary template fits, including component names, near-infrared spectral types and relative $JHK$ magnitudes on the MKO\footnote{Mauna Kea Observatory filter set; see \citet{2002PASP..114..180T} and \citet{2002PASP..114..169S}.} photometric system.  
Averaging over all fits with an F-test statistical weighting, we infer mean component types of M8.5$\pm$0.5 and T5$\pm$1, both rounded off to the nearest half-subclass.  The near-infrared classification of the primary is consistent with the combined-light optical classification.    
We also estimate relative magnitudes of 3.2$\pm$0.3, 3.8$\pm$0.5 and 4.4$\pm$0.6 in the $J$-, $H$- and $K$-bands.  The component properties of {\namesh}
are essentially identical to those inferred for 2MASS~J0320$-$0446AB by \citet{2008ApJ...681..579B}, which is not unexpected given the similarity in their near-infrared spectra (Figure~\ref{fig:spexspec}).

\section{Radial Velocity Variability and Orbit}

\subsection{Radial Velocity Measurements}

While our spectral template analysis provides compelling evidence that {\namesh} harbors a T dwarf companion, we sought independent verification by searching for radial velocity variations in the NIRSPEC data.  Velocities were determined by cross-correlating the {\namesh} and LP~704-48 spectra with each of the radial velocity templates observed in the same order using the IDL {\it xcorl} package \citep{2003ApJ...583..451M,2009ApJ...693.1283W}.  We correlated the spectra over the wavelength ranges 1.283--1.299~$\micron$ (order 59), 1.5455--1.5665~$\micron$ (order 49) and 2.297--2.325~$\micron$ (order 33), in each case sampling ten independent windows equally-spaced in logarithmic wavelength.  The individual correlations within a given order were combined using an outlier-resistant mean based on comparison of median and median absolute deviation statistics (with a 2.5$\sigma$ threshhold), and velocities determined from multiple standards in a given epoch (Table~\ref{tab:rvmeasures}) were combined using an uncertainty-weighted mean.  
We also measured the radial velocities for each of the standards in order to assess systematic effects.  We found these measurements, with uncertainties ranging from 0.5--0.9~{\kms}, were all consistent, albeit with lower precision, with those reported in \citet{2010ApJ...723..684B} to within 1.6$\sigma$.  While a forward-modeling approach would have likely produced better precision, our uncertainties are nevertheless roughly equivalent to contemporary NIRSPEC RV measurements of very cool dwarfs (e.g., \citealt{2002ApJ...569..863P,2007ApJ...666.1205Z,2008ApJ...678L.125B,2010ApJ...723..684B,2010ApJS..186...63R,2012A&A...538A.141R}).

\begin{deluxetable*}{lcccccccccc}
\tablecaption{Radial Velocity Measurements of {\namesh} and LP~704-48 by Comparator (in {\kms})}
\label{tab:rvmeasures}
%\rotate
\tabletypesize{\tiny}
\tablewidth{0pt}
\tablehead{
\colhead{MJD}	&	\colhead{Order}	&	\colhead{J0921-2104}			&	\colhead{J0523-1403}			&	\colhead{J1515+4849}			&	\colhead{J1731+2721}			&	\colhead{J2224-0158}			&	\colhead{J1022+5825}			&	\colhead{J1705-0516}			&	\colhead{BRI 0021-0214}			&	\colhead{LHS 2065}			\\
	&		&	(80.53	$\pm$	0.11)	&	(12.21	$\pm$	0.09)	&	($-$29.97	$\pm$	0.14)	&	($-$29.76	$\pm$	0.11)	&	($-$37.55	$\pm$	0.09)	&	(19.29	$\pm$	0.11)	&	(12.19	$\pm$	0.11)	&	(4.4	$\pm$	2.2)	&	(8.7	$\pm$	1.5)	\\
}
\startdata
\multicolumn{11}{c}{\namesh} \\
\cline{1-11}
55427.53279	&	59	&	-7.36	$\pm$	0.38	&	-7.08	$\pm$	0.46	&	\nodata			&	\nodata			&	\nodata			&	\nodata			&	\nodata			&	\nodata			&	\nodata			\\
55526.27187	&	59	&	-18.39	$\pm$	0.50	&	-17.99	$\pm$	0.36	&	\nodata			&	\nodata			&	\nodata			&	\nodata			&	\nodata			&	\nodata			&	\nodata			\\
55748.61963	&	59	&	-15.57	$\pm$	0.93	&	-15.4	$\pm$	1.5	&	\nodata			&	\nodata			&	\nodata			&	\nodata			&	\nodata			&	\nodata			&	\nodata			\\
55811.54273	&	33	&	\nodata			&	\nodata			&	-20.90	$\pm$	0.65	&	-20.22	$\pm$	0.28	&	-21.1	$\pm$	1.1	&	-22.17	$\pm$	0.38	&	-22.8	$\pm$	1.1	&	\nodata			&	\nodata			\\
55814.53006	&	59	&	-20.17	$\pm$	0.74	&	-19.07	$\pm$	0.85	&	\nodata			&	\nodata			&	\nodata			&	\nodata			&	\nodata			&	\nodata			&	\nodata			\\
55932.23208	&	33	&	\nodata			&	\nodata			&	-22.60	$\pm$	0.77	&	-22.89	$\pm$	0.45	&	-23.3	$\pm$	1.4	&	-24.32	$\pm$	0.51	&	-25.35	$\pm$	0.45	&	\nodata			&	\nodata			\\
55941.19343	&	49	&	\nodata			&	\nodata			&	\nodata			&	\nodata			&	\nodata			&	\nodata			&	\nodata			&	-23.3	$\pm$	2.4	&	-20.9	$\pm$	1.6	\\
\cline{1-11}
\multicolumn{11}{c}{LP~704-48} \\
\cline{1-11}
55932.23208	&	33	&	\nodata			&	\nodata			&	-13.84	$\pm$	0.63	&	-14.39	$\pm$	0.67	&	-13.6	$\pm$	1.2	&	-16.72	$\pm$	0.33	&	-15.10	$\pm$	0.64	&	\nodata			&	\nodata			\\
\enddata
\tablecomments{Radial velocities (RVs) are given in the heliocentric reference frame for the observed Modified Julian Date (MJD; Julian Date - 2400000).}
\end{deluxetable*}

\begin{deluxetable}{llccc}
\tablecaption{Radial Velocity Measurements for {\namesh}}
\label{tab:rvs}
\tabletypesize{\scriptsize}
\tablewidth{0pt}
\tablehead{
\colhead{Date (UT)} &
\colhead{MJD} &
\colhead{Order} &
\colhead{RV} & 
\colhead{$\sigma$} \\ 
&  & & \colhead{(\kms)} & \colhead{(\kms)} \\
\\
}
\startdata
2010 Aug 19 & 55427.53279 & 59 & $-$7.25 & 0.30 \\
2010 Nov 26 & 55526.27187 & 59 & $-$18.13 & 0.30 \\
2011 Jul 06 & 55748.61963 & 59 & $-$15.51 & 0.78 \\
2011 Sep 07 & 55811.54273 & 33 & $-$20.98 & 0.92 \\
2011 Sep 10 & 55814.53006 & 59 & $-$19.70 & 0.56 \\
2012 Jan 06 & 55932.23208 & 33 & $-$24.0 & 1.1 \\
2012 Jan 15 & 55941.19343 & 49 & $-$21.6 & 1.3 \\
\enddata
\tablecomments{Radial velocities (RVs) are given in the heliocentric reference frame for the observed Modified Julian Date (MJD; Julian Date - 2400000).}
\end{deluxetable}

Table~\ref{tab:rvs} lists the final radial velocity measurements for {\namesh} for each epoch. There is clear variation in the velocities, with values ranging from $-$24 to $-$7~{\kms} over the 16 months of observation.  These values straddle the radial velocity measured for LP~704-48, $-$15.6$\pm$0.4~{\kms}, which we consider to be an estimate of the systemic radial velocity (see below).  The $\chi^2$ for these data, 1000 with 6 degrees of freedom, rules out a constant radial motion, indicating the presence of a gravitationally perturbing companion.  

\subsection{Radial Velocity Orbit: MCMC Analysis}

With seven measurements in hand, we have a minimum set necessary to constrain a radial velocity orbit for {\namesh}.  To do this, we performed a Monte Carlo Markov Chain (MCMC) analysis following \citet{2005AJ....129.1706F}.   
We defined the parameter vector:
\begin{equation}
\vec{{\theta}} = \{\ln{P},\ln{K_1},\sqrt{e}\cos{\omega},\sqrt{e}\sin{\omega},\phi_1,V_{com} \}, 
\end{equation}
where $P$ is the period of the orbit in years, $K_1$ the semi-amplitude of the primary's orbital motion along the line of sight, $e$ the eccentricity, $\omega$ the argument of periastron, $\phi_1$ the mean anomaly in the first epoch ($T_1$) of data, and $V_{com}$ the system's center of mass radial velocity.
Note that the epoch of periastron passage is $T_0$ = $T_1 - P{\phi_1}/{2\pi}$.
As discussed in  \citet{2005AJ....129.1706F}, this particular configuration of orbital parameters is found to improve convergence for  MCMC analysis in cases of small $e$.
The semi-amplitude, period and eccentricity are related to the mass function:
\begin{equation}
f_M\sin{i} \equiv  \frac{{\rm M}_2\sin{i}}{({\rm M}_1+{\rm M}_2)^{2/3}} = K_1\left(\frac{P}{2\pi{G}}\right)^{1/3}\sqrt{1-e^2}
\label{eqn:fm}
\end{equation} 
where $i$ is the inclination of the orbital plane with respect to the plane of the sky (0$\degr$ corresponds to a face-on projection), and M$_1$ and M$_2$ are the masses of the M and T dwarf components of {\namesh}, respectively.
We performed three sets of 300 MCMC trials to maximally explore the six-dimensional parameter space.  For each trial, we started with an initial 
parameter set $\vec{\theta}_0$ selected from distributions derived from the observational data.  For  $P$ and $K_1$, we drew from Gaussian distributions with means of 0.5~yr and 9~{\kms} and widths  0.2~yr and 3~{\kms}, derived from the scales over which the radial velocities reversed in trend.  
For $e$, $\omega$ and $\phi_1$, we drew from uniform distributions spanning ranges [0,0.9], [0,2$\pi$] and [0,2$\pi$], respectively.  For $V_{com}$, we drew from a Gaussian distribution centered on the observed radial velocity of LP~704-48 with a standard deviation of 0.6~{\kms}, under the assumption that the relative orbital motion of the pair is negligible ($\lesssim$0.5~{\kms})
Throughout the simulation, we imposed boundary conditions on the parameters of 0.05 $\lesssim P \lesssim$ 10~yr, 1 $\lesssim K_1 \lesssim$ 100~{\kms} and $e \leq$ 0.95.  

For each step in the MCMC chain, we computed a new vector $\vec{\theta}^{\prime}$ by
changing one parameter $\theta_j$, drawing from a Gaussian distribution
\begin{equation}
q(\theta_j^{\prime}|\theta_j) \propto e^{-\frac{(\theta_j^{\prime}-\theta_j)^2}{2\beta_j}},
\end{equation}
where $\vec{\beta}$ is a vector of scale factors.  Initial scale factors were chosen by trial and error to optimize acceptance rates \citep{doi:10.1214/ss/1177011136}.  We also found it useful to allow these scale factors to increase by 20\% (up to a factor of 20) every 500 chain steps after the associated parameter was last updated.  With each parameter set, we computed that model's radial velocities (${V}_{mod}$) at the epochs of observation and the $\chi^2$ deviation with the measured values (${V}_{obs}$):
\begin{equation}
\chi^2 = \sum_k\left(\frac{{V}_{mod,k}-{V}_{obs,k}}{\sigma_{obs,k}}\right)^2 + \left(\frac{V_{com}-V_{LP~704-48}}{\sigma_{LP~704-48}}\right)^2,
\end{equation}
where the sum is over all observations.  Note that we have explicitly included the measurement of the radial motion of LP~704-48 as an ``observation'' of the systemic motion for {\namesh}AB.
%We performed a second series of MCMC trials that dropped this constraint.
The new parameter set $\vec{\theta}^{\prime}$ replaced $\vec{\theta}$ if
\begin{equation}
U(0,1) \leq e^{({\chi^2}^{\prime}-{\chi^2})/2},
\end{equation}
where $U(0,1)$ is a random number drawn from a uniform distribution between 0 and 1. This process was repeated cyclically for each parameter 100,000 times for each of the 300 trials.

\begin{figure}
\plotone{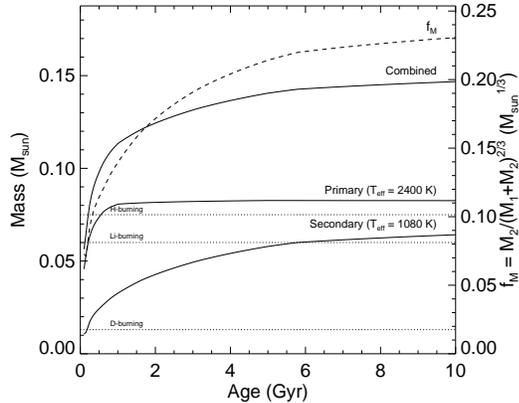}
\caption{Expected component and combined masses of {\namesh}AB (solid black curves) as a function of age based on the evolutionary models of \citet{2003A&A...402..701B} and assuming component {\teff}s of 2400~K and 1080~K for the M8.5 primary and T5 secondary, respectively.  The system mass function, $f_{\rm M} = {\rm M}_2/({\rm M}_1+{\rm M}_2)^{2/3}$ (dashed line)  is also shown, with values corresponding to the right axis.  Hydrogen-, lithium- and deuterium-burning mass limits are indicated by the horizontal dotted lines.
%, while the estimated minimum age of the LP~704-48 primary (based on its lack of H$\alpha$ emission) is indicated by the vertical dotted line.
\label{fig:model}}
\end{figure}

While these data are nominally sufficient to constrain the radial velocity orbit of the system, we also incorporated constraints based on the physical properties of the components as inferred from their spectral types and evolutionary models. Using the 
\citet{2009ApJ...702..154S} {\teff}/SpT relation, we adopted component effective temperatures of 2400$\pm$160~K and 1080$\pm$170~K for {\namesh}A and B, respectively, and used these to estimate component masses as a function of age with the evolutionary models of \citet{2003A&A...402..701B}.  Figure~\ref{fig:model} plots these masses, the combined system mass and the mass function (Eqn.~\ref{eqn:fm}) for ages of 0.1--10~Gyr.  Note that the estimated mass of {\namesh}A asymptotes to 0.082~{\msun} beyond 2~Gyr, while the estimated mass of {\namesh}B (and by extension the combined mass and mass ratio) increase monotonically from 0.033~{\msun} to 0.064~{\msun} for ages 1--10~Gyr, never breaching the minimum mass for hydrogen burning. The mass ratio ($q \equiv$ M$_2$/M$_1$) ranges from 0.41 at 1~Gyr to 0.78 at 10~Gyr.
The evolutionary estimates of $f_M$ constrain the allowed orbital solutions by the requirement that 
\begin{equation}
f_M^{orbit}\sin{i} \leq {\rm maximum}(f_M^{model})\approx 0.23~{\rm M}_{\odot}^{1/3}
\label{eq:fmmax}
\end{equation}
(i.e., $\sin{i} \leq 1$). 
To explore the systematic effects of our external constraints on the orbit model, we performed separate MCMC trials with both $V_{COM}$ and evolutionary model constraints (Case A), without  the $V_{COM}$ constraints (Case B) and with neither the $V_{COM}$ nor evolutionary model constraints (Case C).

\subsection{Results}

\begin{figure}
\centering
\epsscale{1}
\plotone{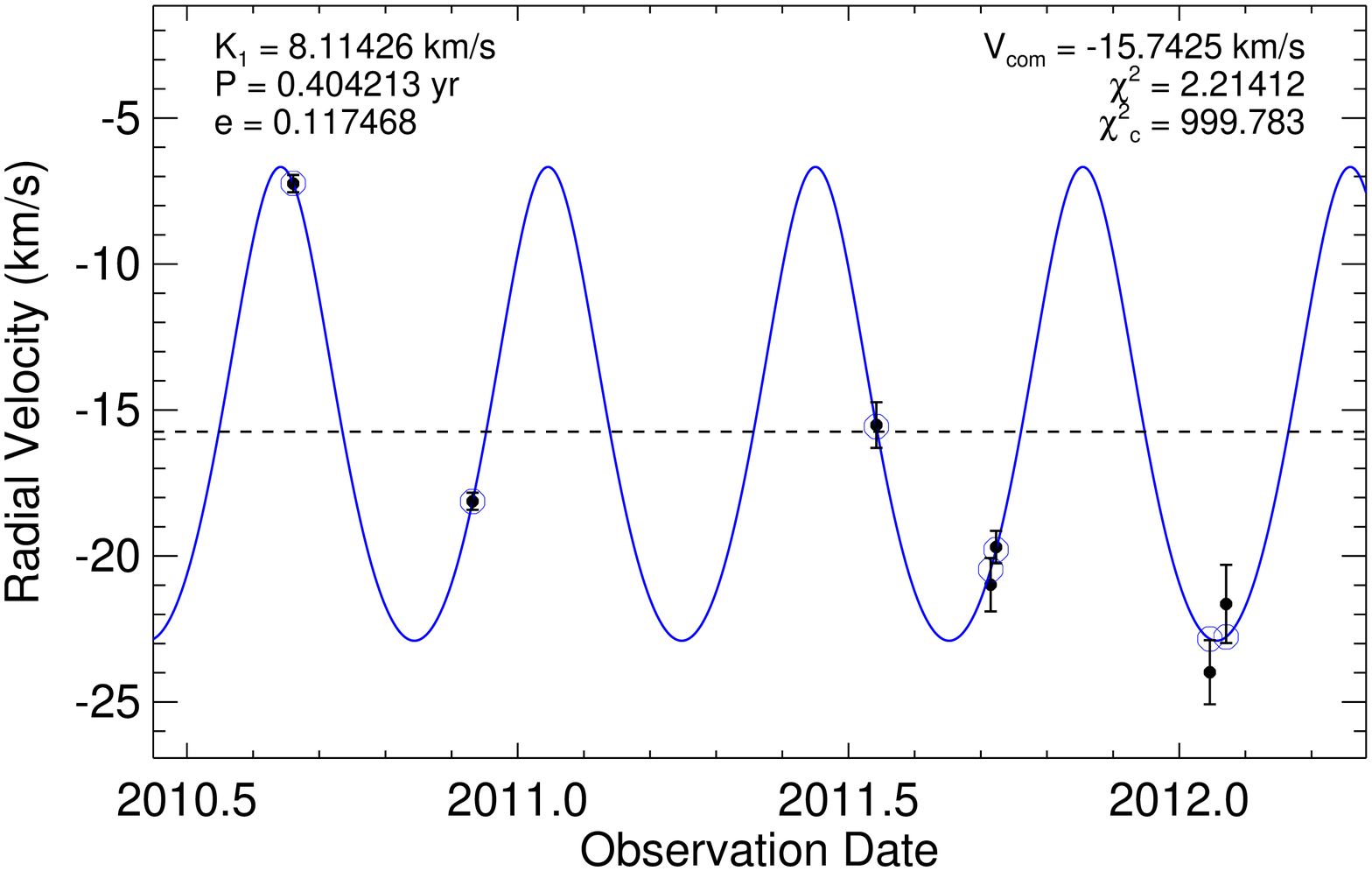} 
\plotone{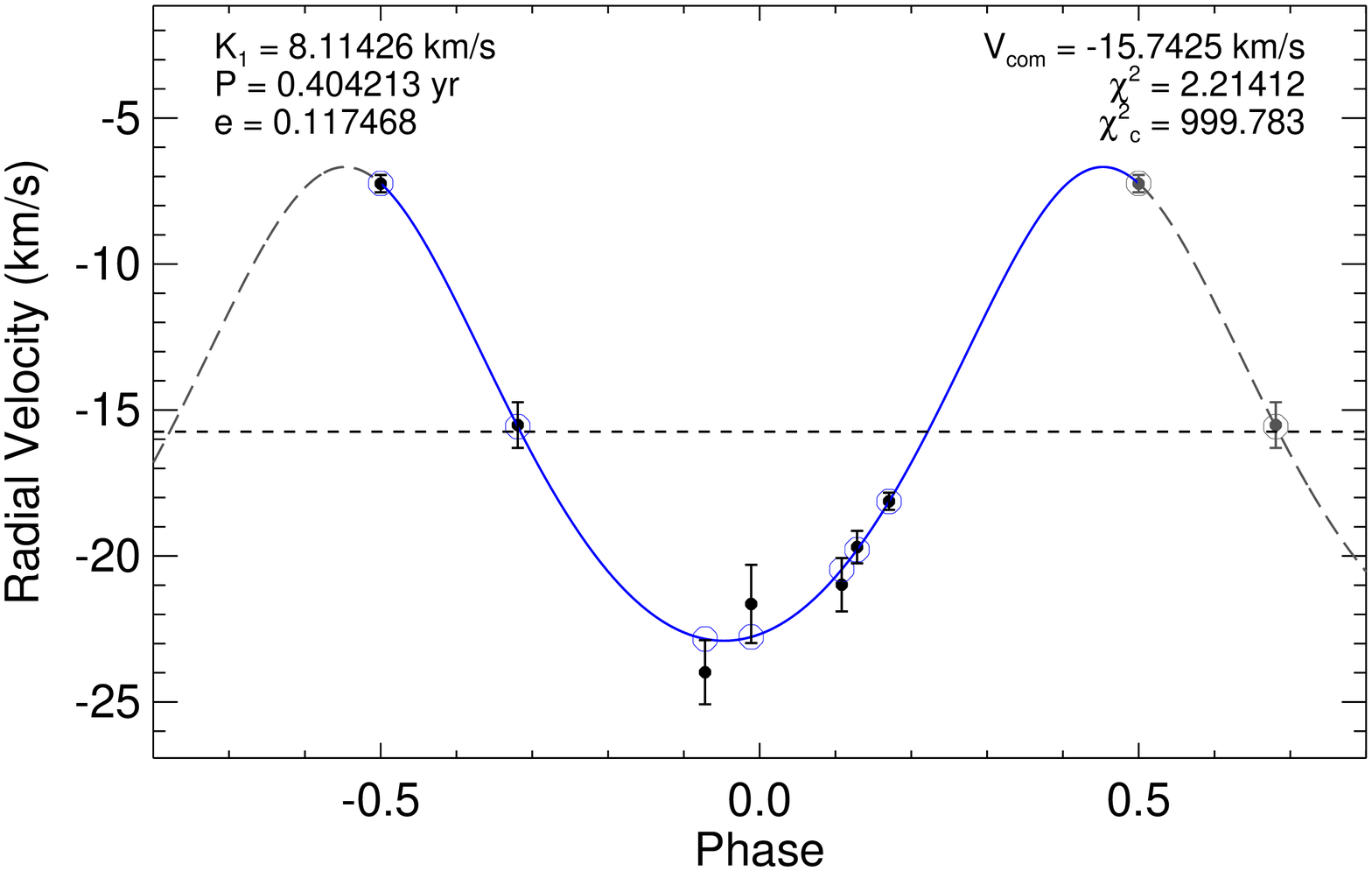}
\caption{Best-fit primary radial velocity orbit solution for {\namesh}AB based on our Case A MCMC fit.  The top panel shows the full radial velocity curve with measurements (solid circles with error bars) as a function of observation date; the bottom panel shows the velocity curve as a function of phase.  Open circles indicate the expected radial velocity at a given observational epoch based on the best-fit model (blue lines).  Orbital parameters $K_1$, $P$, $e$, and $V_{COM}$ (horizontal dashed line) are listed, as is the $\chi^2$ value for the fit and for a constant velocity solution ($\chi^2_c$).
\label{fig:orbits}}
\end{figure}

\begin{figure}
\centering
\epsscale{1}
\plotone{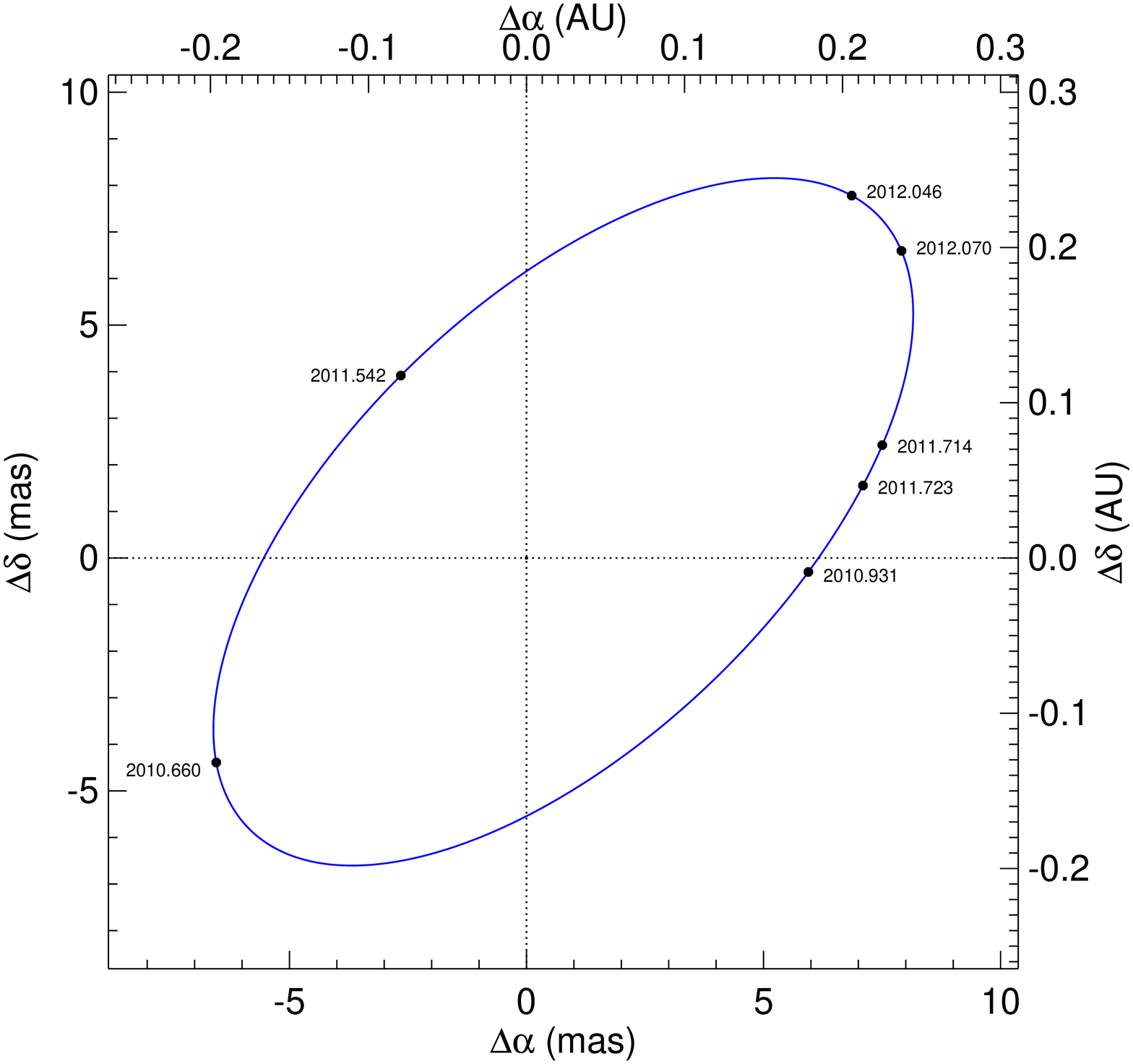} 
\caption{Predicted relative astrometric orbit of the LP~704-48/{\namesh}AB system based on the best-fit radial velocity orbit, and assuming $i$ = $i_{min}$ and an arbitrary longitude of nodes.  The primary is located at the origin.  Both apparent (bottom and left axes) and projected (top and right axes) coordinates are shown, and epochs at which radial velocity measurements were obtained are labeled. 
\label{fig:astrometry}}
\end{figure}

Figure~\ref{fig:orbits} displays the best-fit orbital solution for our Case A MCMC analysis. 
The observed radial velocities are well-fit to a low-eccentricity ($e$ = 0.12), short-period (0.404~yr = 148~days) orbit, with the first and last observations taken around maximum elongation.
The inferred center-of-mass velocity, $-$15.7~{\kms}, is consistent with the measured velocity
of LP~704-48, although this is not surprising for our Case A analysis given the $V_{COM}$ constraint.    
With a $\chi^2$ = 2.21, this solution is an excellent fit to the data, and suggests that the uncertainties on the radial velocities may even be slightly overestimated.   
The best case solutions for our Case B and Case C analyses have similar parameters, notably yielding the same $V_{COM}$ values without a constraint, and comparable $\chi^2$ values.
Figure~\ref{fig:astrometry} displays a representative relative astrometric orbit for the best-fit Case A 
solution.
The apparent separation of the two components based on the solution never exceeds 10~mas, making
{\namesh}AB unresolvable with current AO or space-based instrumentation.

Marginalized individual and joint probability distributions for various orbit parameters and parameter pairs 
were derived by first computing a weight $\wp$ for each solution $i$ in the chain based on its $\chi^2$ value relative to the minimum $\chi^2$ of the chain (i.e., the best-fit solution):
\begin{equation}
\wp_i = e^{-({\chi^2}_i-{\rm min}[\{\chi^2\}])/2}.
\label{eqn:significance}
\end{equation}
For computational ease, we excluded all parameter sets with $\wp < 10^{-4}$, which left of order 10$^5$ solutions per Case\footnote{This pruning is equivalent to the common practice of excluding some fraction of the initial steps in a given MCMC chain; see for example \citet{2005AJ....129.1706F}.}.  For individual parameters, we divided the full range of parameter values $\{\theta_j\}$ into 30 bins, and for each bin summed the probabilities of all solutions with $\theta_j$ in that bin.  The resulting marginalized probability distributions were normalized and then fit with Gaussians to determine parameter means and standard deviations.  For parameter pairs, we performed the same analysis, dividing the ranges of both parameters into 30 bins and summing the probabilities of those solutions whose parameters simultaneously satisfy both bin ranges. 

\begin{figure}
\epsscale{1.0}
\centering
\plotone{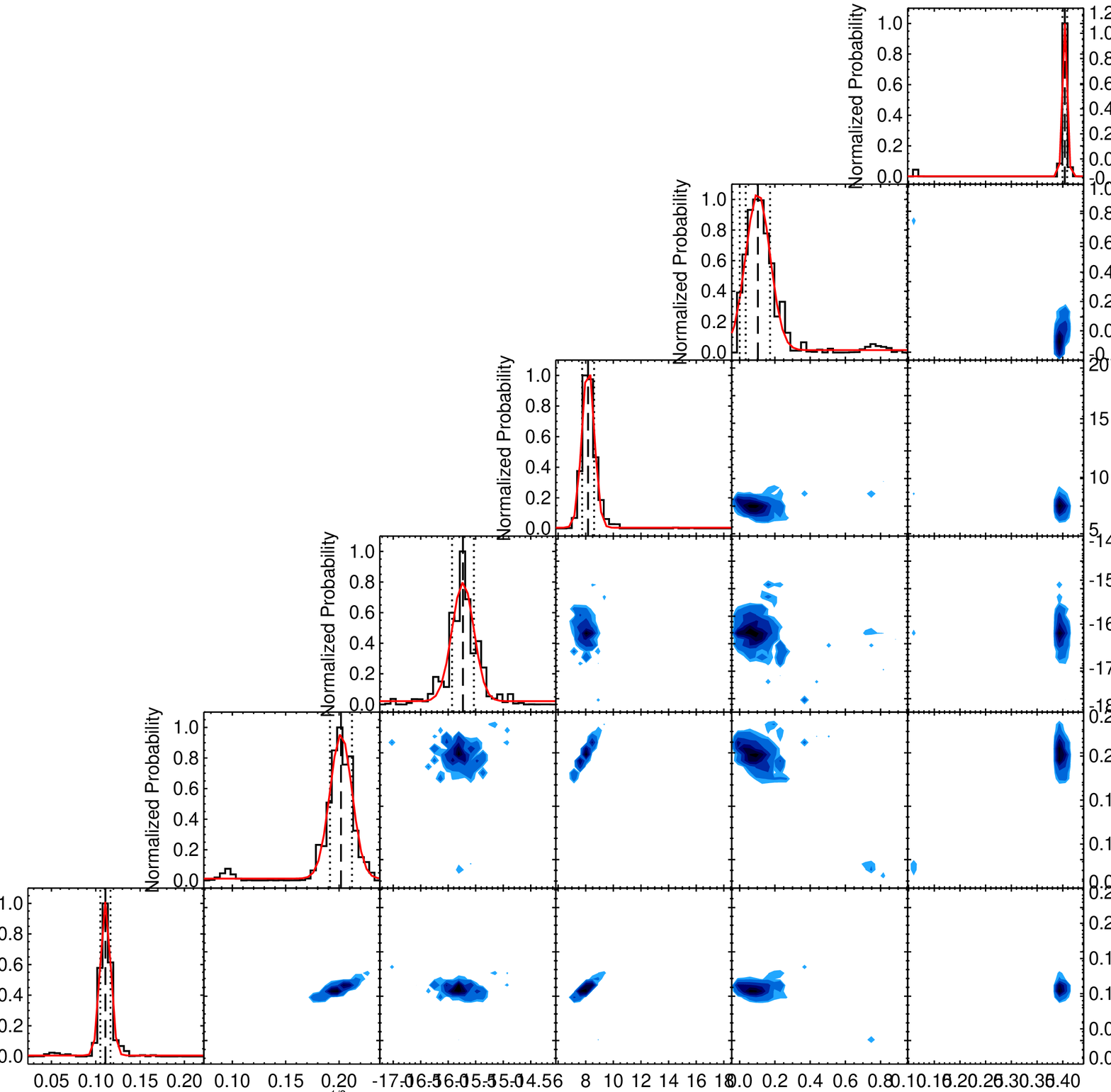} \\ ~\\
\caption{Probability distributions of derived orbit parameters
for {\namesh}AB based on our Case A MCMC fit. The leftmost panels
show the marginalized probability distributions of (from top to bottom)
period (in yr), eccentricity, primary radial velocity semi-amplitude ($K_1$ in {\kms}),
center of mass velocity ($V_{COM}$ in {\kms}), 
mass function ($f_M$ = M$_2\sin{i}/$(M$_1$+M$_2$)$^{2/3}$ in M$_{\sun}^{1/3}$),
and primary semimajor axis ($a_1\sin{i}$ in AU).
Gaussian fits to these distributions used to determine mean values
and standard deviations are shown in red.
The other panels display joint probabilities distributions between
parameter pairs, which shading from light to dark blue tracing factors
of  0.05, 0.1, 0.3, 0.5, 0.7 and 0.9 relative to the best-fit solution (see Eqn.~\ref{eqn:significance}).  
\label{fig:orbit_distributions}}
\end{figure}

\begin{deluxetable}{lcccc}
\tablecaption{Inferred Orbital Parameters for {\namesh}AB \label{tab:orbit}}
\tabletypesize{\scriptsize}
\tablewidth{0pt}
\tablehead{
\colhead{Parameter} &
\colhead{\bf Case A} &
\colhead{Case B} &
\colhead{Case C} &
 \\
}
\startdata
Minimum $\chi^2$ & 2.21 & 2.38 & 1.98 \\
Period (yr) & 0.404$\pm$0.004 & 0.407$\pm$0.009 & 0.407$\pm$0.007 \\
Period (dy) & 147.6$\pm$1.5 & 148.7$\pm$3.3 & 148.7$\pm$2.6 \\
Eccentricity & 0.10$\pm$0.07 & 0.22$\pm$0.16 & 0.10$\pm$0.06 \\
$\omega$ ($\degr$) & $-0.3{\pm}2.1$ & $-67{\pm}4$ & $-0.1{\pm}2.2$  \\
T$_0$ (MJD) & 55421$\pm$3 & 55421$\pm$5 & 55421$\pm$4 \\
$K_1$ ({\kms}) & 8.2$\pm$0.4 & 8.5$\pm$0.6 & 8.4$\pm$0.6 \\
$V_{COM}$ ({\kms}) & $-$15.7$\pm$0.2 & $-$16.4$\pm$0.7 & $-$15.9$\pm$0.5 \\
$f_M$ (M$_{\sun}^{1/3}$) & 0.202$\pm$0.010 & 0.205$\pm$0.013 & 0.202$\pm$0.010 \\
$a_1\sin{i}$ (AU)  & 0.111$\pm$0.005 & 0.122$\pm$0.011 & 0.112$\pm$0.018 \\
$a$ (AU)\tablenotemark{a} & 0.286$\pm$0.009 & 0.288$\pm$0.010 & \nodata \\
Minimum $i$ ($\degr$)\tablenotemark{a} & 61$\pm$5 & 62$\pm$6 & \nodata \\
Minimum Age (Gyr)\tablenotemark{a} & 3.7$\pm$0.8 & 4.0$\pm$1.0 & \nodata \\
\enddata
\tablenotetext{a}{Assuming a combined system mass of 0.144$\pm$0.013~{\msun} based on the evolutionary models of Burrows et al.~(2001) and Baraffe et al.~2003.}
\tablecomments{Case A incorporates constraints on both $V_{COM}$ and $f_M$; Case B drops the $V_{COM}$ constraints; Case C drops both $V_{COM}$ and $f_M$ constraints; see Section~5.2.}
\end{deluxetable}

As summarized in Table~\ref{tab:orbit}, we derive fairly stringent ($\lesssim$10\%) constraints on the 
orbital period (0.404$\pm$0.004~yr, 147.6$\pm$1.5~days),
eccentricity (0.10$\pm$0.07),
and $K_1$ (8.2$\pm$0.4~{\kms}), and determine 
$\omega$ to within 2$\degr$ and periapse passage $T_0$ to within 3 days (periapse occurred
6 days prior to our first observation).  
These parameters yield
$f_M\sin{i}$ = 0.202$\pm$0.010~M$_{\sun}^{1/3}$, which is 
near the maximum $f_M$ allowed from the evolutionary models (Eqn~\ref{eq:fmmax}).
Our determination of $V_{COM}$ ($-15.7{\pm}0.2$~{\kms}) is consistent with, and more accurate than, 
the measured radial velocity of LP~704-48.
Inferred parameters for our Case B and Case C analyses were 
identical to within the uncertainties.
We also derived the primary semi-major axis of the system 
\begin{equation}
a_1\sin{i} = \frac{PK_1}{2\pi\sqrt{1-e^2}} = 0.111{\pm}0.005~{\rm AU}
\end{equation}
for Case A, with the other Cases giving statistically equivalent results.
For comparison, the orbit of 2MASS~J0320$-$0446AB, which is also nearly circular,
has a period 1.7 times longer and a 
primary semi-major axis 1.4 times wider than {\namesh}AB \citep{2010ApJ...723..684B}.
Our marginalized parameter pair distributions show that $P$, $e$ and $K_1$ 
values are uncorrelated,
while $f_M\sin{i}$ and $a_1\sin{i}$ values show an expected correlation with $K_1$.

\begin{figure}
\epsscale{0.9}
\centering
\plotone{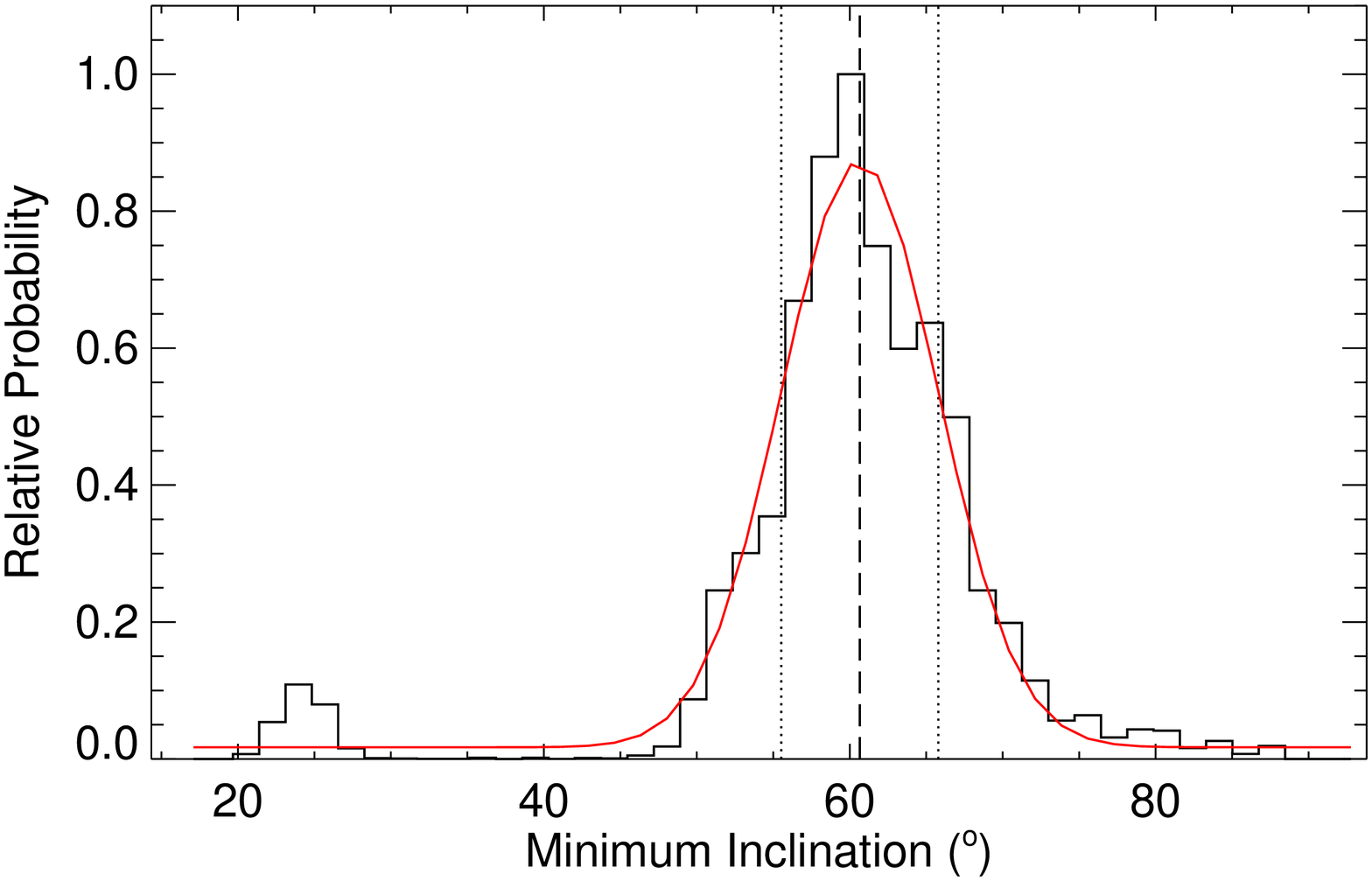}
\plotone{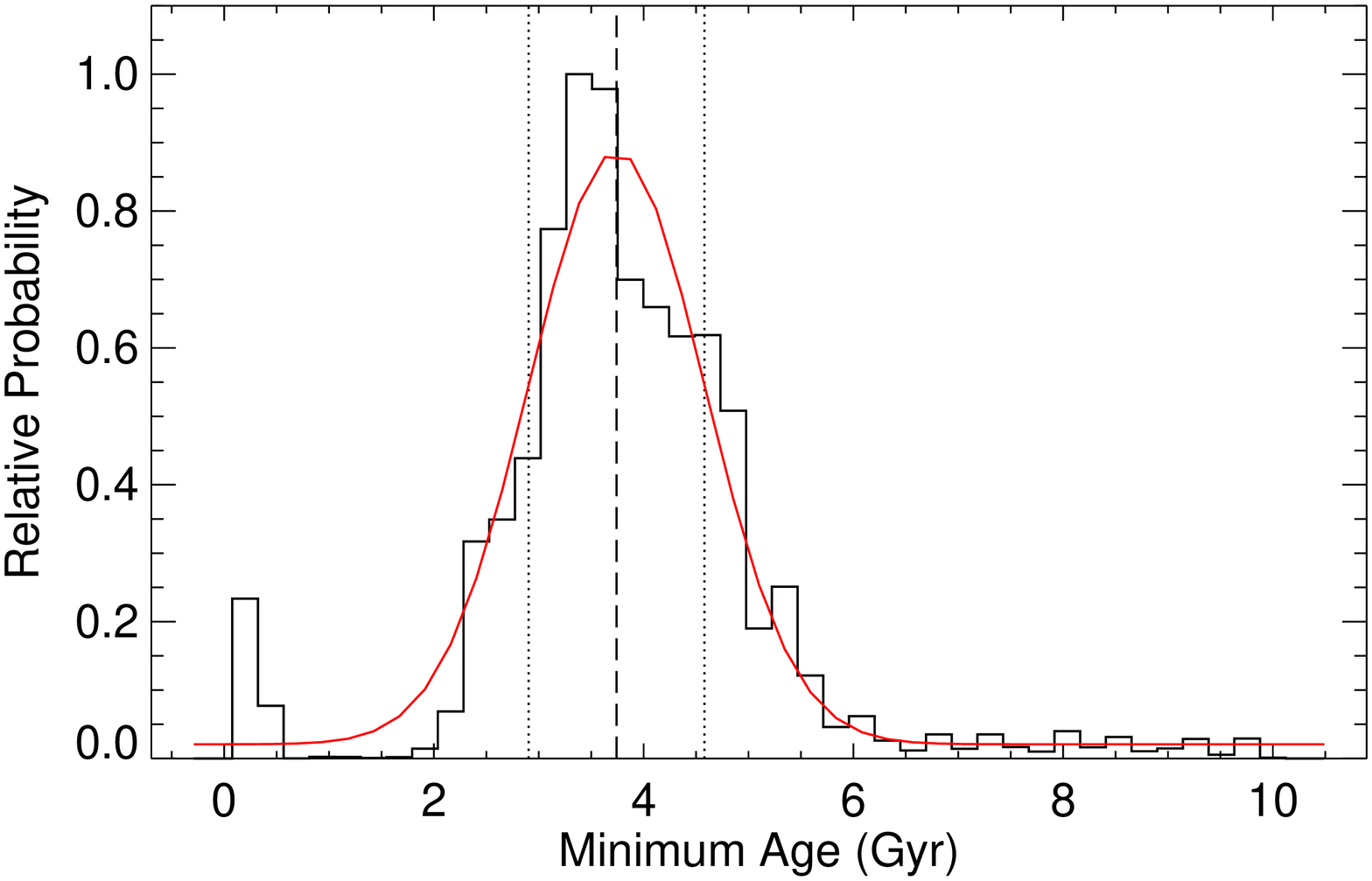}
\caption{Marginalized probability distributions of the minimum orbital inclination (top, in degrees)
and minimum system age (bottom, in Gyr) for {\namesh}AB based on our Case A MCMC fit.
These assume component masses inferred from the spectral types and evolutionary models of \citet{2003A&A...402..701B} (see Figure~\ref{fig:model}).
\label{fig:minincage}}
\end{figure}

For our analyses that included evolutionary mass constraints, Cases A and B, we
determined model-dependent minimum inclinations and system ages (Figure~\ref{fig:minincage}).
From Eqn~\ref{eq:fmmax}, we 
derived $i_{min}$ = 61$\degr$$\pm$5$\degr$ (Case A), 
with solutions up to 70$\degr$ being significant.
However, eclipses remain unlikely for this pair, requiring an inclination $i > 89{\fdg}8.$\footnote{The geometric probability that one component will eclipse is $\frac{R_1+R_2}{a}\frac{1+2e/\pi}{1-e^2} \approx$ 0.4\%, assuming $R_1 = R_2 = R_{Jupiter}$.}
Constraints on system age is also based on the requirement that $\sin{i} \leq 1$, which implies a minimum value for the time-dependent $f_M$ (Figure~\ref{fig:model}).
Our marginalized probability distribution for the minimum age for Case A peaks at 3.7$\pm$0.8~Gyr, consistent with the 8$^{+0.5}_{-1.0}$~Gyr minimum age inferred for LP~704-48 based on its lack of H$\alpha$ emission.  
Formally adopting a system age of 3--10~Gyr limits the mass of the secondary to 0.049--0.064~{\msun},
the total system mass to 0.131--0.147~{\msun},
and the mass ratio of the components to 0.60 $\leq q \leq$ 0.78.\footnote{This is based on the evolutionary models of \citet{2003A&A...402..701B}. The models of \citet{2008ApJ...689.1327S} yield similar values; the models of \citet{2001RvMP...73..719B} yield M$_2$ = 0.055--0.072~{\msun}, M$_1$+M$_2$ = 0.139--0.157~{\msun}, and $q$ =  0.66--0.85.}.
%{\namesh}AB therefore appears to be an older VLM binary with components that straddle the hydrogen burning minimum mass, with the low {\teff} of the secondary arising from the relatively long cooling time of the system.  
The total system mass can be combined with the period to infer 
a (model-dependent) semimajor axis of 0.286$\pm$0.009~AU.  This makes {\namesh}AB the third tightest VLM binary discovered to date, behind the spectroscopic binary PPL~15AB (\citealt{1999AJ....118.2460B}; $a$ = 0.03~AU) and the 1~Myr eclipsing brown dwarf binary 2MASS~J05352184$-$0546085 (hereafter 2MASS~J0535$-$0546; \citealt{2006Natur.440..311S}; $a$ = 0.04~AU),
and tighter than 2MASS~J0320$-$0446 ($a$ = 0.4~AU; \citet{2010ApJ...723..684B}).

\section{Discussion}

\subsection{Very Low Mass Triples}

The LP~704-48/{\namesh}AB system joins a growing list of VLM triples in which all three components have masses near or below 0.1~{\msun} (assuming a mass of 0.093~{\msun} for LP~704-48 based on the evolutionary models used above).  
Formally defining the category ``VLM triple'' as a bound system of three hydrogen-rich objects with a total mass of 0.3~{\msun} or less, we list the seven candidate and confirmed systems that fall into this category
in Table~\ref{tab:triples}.
With the exception of LHS~1070ABC, these systems are strongly hierarchical, with a ratio of outer to inner
separations ranging from $\sim$5 for LP~714-37ABC to $\sim$2900 for LP~704-48/{\namesh}AB.
They also have estimated component masses that are comparable to each other, although this may be a selection bias (sensitivity limits) and subject to the typically large uncertainties in mass estimates for substellar dwarfs.
Three of these systems have total estimated masses M$_{tot}$ $\lesssim$ 0.15~{\msun}, although their tertiary components have not been solidly confirmed.
DENIS-P~J020529.0-115925ABC (hereafter DENIS~J0205-1159AB(C); \citealt{1997A&A...327L..25D,1999ApJ...526L..25K,2005AJ....129..511B}) was resolved into a $\sim$7~AU binary by AO and HST imaging. Evidence for a third component was reported by \citet{2005AJ....129..511B}, but only on the basis of persistent residuals in point spread function fitting of the resolved pair in multi-epoch HST images.
2MASS J08503593+1057156ABC (hereafter 2MASS~J0850+1057AB(C); \citealt{1999ApJ...519..802K,2001AJ....121..489R}) was resolved as a 0$\farcs$16 binary with HST, and subsequently identified as a candidate triple based on the later classification
of its primary, despite this component being roughly a magnitude brighter in the near-infrared \citep{2011AJ....141...70B}. This result has been called into question by \citet{2012arXiv1201.2465D}, however, based on reanalysis of the combined-light spectrum.
Kelu~1AB(C) \citep{1997ApJ...491L.107R,1999Sci...283.1718M,2005ApJ...634..616L,2006PASP..118..611G} was resolved as a 6~AU visual double with AO and HST imaging, and its  primary identified as an unresolved ($<$4~AU) L/T spectral binary by \citet{2008arXiv0811.0556S}, although there has been no evidence of RV variability in the combined-light system at the 1--3~{\kms} level \citep{2006AJ....132..663B,2010ApJ...723..684B}. 
Given the uncertain nature of these triple candidates, we argue that LP~704-48/{\namesh}AB is currently the lowest-mass triple verified through multiple techniques.  It is also the only VLM triple system in Table~\ref{tab:triples} for which multiple orbits of the tight inner binary have been observed. 

\begin{deluxetable*}{lllllccccccl}
\tablecaption{Confirmed and Candidate Very Low Mass Triples (M$_{tot} \lesssim 0.3$~{\msun}) \label{tab:triples}}
\tabletypesize{\scriptsize}
\tablewidth{0pt}
\tablehead{
\multicolumn{2}{c}{Components\tablenotemark{a}}  & 
\multicolumn{3}{c}{Spectral Types}  & 
\multicolumn{4}{c}{Estimated Masses ({\msun})} &
\multicolumn{2}{c}{Separations (AU)}  & 
\colhead{Ref} \\
%\cline{1-2} \cline{3-5} \cline{6-9} \cline {10-11}
\colhead{A} &
\colhead{BC} &
\colhead{A} & \colhead{B} & \colhead{C} & 
\colhead{A} & \colhead{B} & \colhead{C} &  \colhead{Total} &
\colhead{A-BC} & \colhead{BC} & \\
}
\startdata
LP~714-37A & LP~714-37BC & M5.5 & M8 & M8.5 & 0.11 & 0.09 & 0.08 & 0.28 & 36$\pm$5 & 6.8$\pm$0.9 & 1 \\
LHS~1070A & LHS~1070BC & M5.5 & M8.5 & M9 & 0.115  &  0.08 & 0.077 & 0.27 & $\sim$12\tablenotemark{b} &  3.57$\pm$0.07 & 2,3 \\
LP~213-68 & LP~213-67AB & M8 & M8 & L0 & 0.09  & 0.09 & 0.084 & 0.27 & 340$\pm$60 & 2.9$\pm$0.6 & 4,5 \\
{\bf LP~704-48} & {\bf SDSS~J0006$-$0852AB} & M7 & M8.5 & T5 & 0.092 & 0.083 & 0.056 & 0.23 & 820$\pm$120 & 0.286$\pm$0.009 & 6 \\
DENIS~J0205-1159A & DENIS~J0205-1159B(C)\tablenotemark{c} &  L5 & L8 & T0: & 0.06 & 0.05 & 0.04 & 0.15 & 7$\pm$1 & $\sim$1.3 & 7 \\
2MASS~J0850+1057B & 2MASS~J0850+1057A(C)\tablenotemark{c} & L6 & L7 & L7: & 0.05 & 0.05 & 0.05 & 0.15 &  6.0$\pm$0.9 & $<$4  & 8 \\
Kelu~1B & Kelu~1A(C)\tablenotemark{c} & L3p & L0.5 & T7: & 0.05 & 0.06 & 0.02 & 0.13 &  6.4$^{+2.4}_{-1.3}$ & $<$4  & 9 \\
%G~124-62 & DENIS~J1441$-$0945AB & M4.5e & L1 & L1 &  & 0.072 & 0.072 & 
\enddata
\tablenotetext{a}{For this table, we refer to the outermost component as ``A'' and the inner binary as ``BC'' irrespective of mass or designation.}
\tablenotetext{b}{Scaling from the inner semi-major axis using the period-mass ratio: $(a_{out}/a_{in})^3 = (P_{out}/P_{in})^2 \times (M_{out}/M_{in})$, and values from Seifahrt et al.~(2008).}
\tablenotetext{c}{Candidate triples with unconfirmed third component.}
\tablerefs{
(1) \citet{2006ApJ...645L.153P}; 
(2) \citet{2001A&A...367..183L}; 
(3) \citet{2008A&A...484..429S}; 
(4) \citet{2000MNRAS.311..385G}; 
(5) \citet{2003ApJ...587..407C}; 
(6) This paper; 
(7) \citet{2005AJ....129..511B}; 
(8) \citet{2011AJ....141...70B};
(9) \citet{2008arXiv0811.0556S}.}
\end{deluxetable*}

The LP~704-48/{\namesh}AB system shares much in common with two recently uncovered, but slightly more massive,
low-mass triples containing transiting substellar components:
NLTT~41135AB/NLTT~41136 \citep{2010ApJ...718.1353I} and 
LHS~6343ABC \citep{2011ApJ...730...79J}.
Both are similarly composed of relatively wide M-dwarf pairs 
(55~AU and 20~AU, respectively) with one component hosting a tightly-orbiting
(0.02~AU and 0.08~AU) substellar mass (0.03~{\msun} and 0.06~{\msun}) companion.
For these systems, the component separations are roughly an order of magnitude smaller than for
LP~704-48/{\namesh}AB, but the relative inner to outer separations and component masses are comparable.  Moreover, 
based on evolutionary models, the tertiaries of these systems are also likely to be T dwarfs.
The fact that three such systems have been identified in the span of three years suggests that such low-mass
triple configurations may be quite common, particularly among wide VLM pairs
(e.g., \citealt{2010ApJ...720.1727L}), although a robust survey is needed to 
quantify the incidence of such systems.

\subsection{Stability of the LP~704-48/{\namesh}AB System}

The LP~704-48/{\namesh}AB system is at an extremum among the VLM triples listed in Table~\ref{tab:triples} in that it has both the widest outer separation and the smallest (measured) inner separation in the sample.  The outer pairing is remarkable for a system with M$_{tot}$ $\lesssim$0.25~{\msun}; currently, only three other VLM field binaries are known to have projected separations $>$500~AU.\footnote{Koenigstuhl-1 at 1800~AU \citep{2007A&A...462L..61C},
2MASS~J0126555$-$502239 at 5100~AU; \citep{2007ApJ...659L..49A,2009ApJ...692..149A}
and 2MASS~J12583501+4013083 at 6700~AU \citep{2009ApJ...698..405R}.}
The gravitational binding energy of LP~704-48 and {\namesh}AB, $|E_b| \lesssim$ (2--3)$\times$10$^{41}$~erg, is low but not unprecedented.  Recently, several wide multiples of comparable total mass and binding energy have been found (e.g., \citealt{2010AJ....139.2566D,2010ApJ...720.1727L}), including high-order systems such as NLTT~20346AB/2MASS~J0850359+105716AB(C) ($|E_b| \approx 0.4\times10^{41}$~erg; \citealt{1999ApJ...519..802K,2001AJ....121..489R,2011AJ....141...70B,2011AJ....141...71F})
and G~124-62/DENIS-P~J144137.3$-$094559AB ($|E_b| \approx 3\times10^{41}$~erg; \citealt{1999AJ....118.2466M,2006A&A...456..253M,2003AJ....126.1526B,2005A&A...440..967S}), both of which contain substellar components.
%\footnote{Note that these systems all reside in the saddle of the bimodal binding energy distribution identified by \citet{2010AJ....139.2566D} among 0.3--1~{\msun} wide binaries, which these authors concluded differentiated young and old populations.} 
In part, it is the additional mass of the T dwarf tertiary that pushes the LP~704-48/{\namesh}AB system into a ``normal'' regime in mass/separation space (see \citealt{2011AJ....141...71F}). This component also contributes to the long-term stability of the wide pair to external perturbation \citep{1987ApJ...312..367W,2010AJ....139.2566D}.

\begin{figure}
\centering
\epsscale{1.1}
\plotone{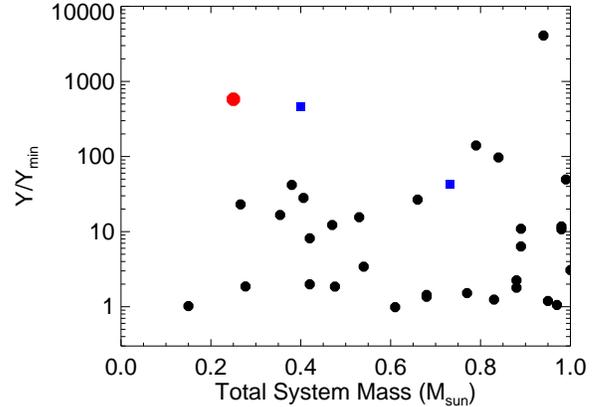} 
\caption{Stability ratio $Y/Y_{min}$ as a function of total system mass for a sample of low mass triple systems drawn from the Multiple Star Catalog \citep{1997A&AS..124...75T}; \citet{2011AJ....141...71F}; and this paper.  The LP~704-48/{\namesh}AB system is indicated by the large red circle at left; the eclipsing triples NLTT~41135AB + NLTT~41136 \citep{2010ApJ...718.1353I} and LHS~6343ABC \citep{2011ApJ...730...79J} are indicated by blue squares.
\label{fig:hierarchical}}
\end{figure}

Because of the large difference in outer to inner separations, 
LP~704-48/{\namesh}AB is also exceptionally stable to internal dynamical disruption.
Internal stability can be quantified with the criteria of \citet{1995ApJ...455..640E},
who examined the ratio of outer periapse $a_{out}(1-e_{out})$ to inner 
apoapse $a_{in}(1+e_{in})$ for triple star systems.  For simplicity, we drop the eccentricity terms and approximate the ratio $Y = a_{out}/a_{in}$, using as $a_{out}$ the projected separation of the wide pairing.  For LP~704-48/{\namesh}AB, $Y \approx 2900$, a value several orders of magnitude greater the critical ratio
\begin{equation}
Y_{min} = 1+\frac{3.7}{q_{out}^{1/3.}} + \frac{2.2}{1+q_{out}^{1/3}} + 1.4q_{in}^{1/3}\frac{q_{out}^{1/3} - 1}{1+q_{out}^{1/3}},
\end{equation}
which is $\approx$5 for this system, assuming $q_{in} \equiv {\rm M}_B/{\rm M}_A \approx 0.7$ for {\namesh}AB and $q_{out} \equiv ({\rm M}_A+{\rm M}_B)/{\rm M}_{LP~704-48} \approx 1.6.$\footnote{These estimates also imply a period ratio $X = Y^{1.5}\left(\frac{q_{out}}{q_{out}+1}\right)^{0.5} \approx 10^5$.}
Figure~\ref{fig:hierarchical} compares the ratio $Y/Y_{min}$ for LP~704-48/{\namesh}AB and other low-mass triples listed in the most recent version (April 2010) of the Multiple Star Catalog \citep{1997A&AS..124...75T}, \citet{2011AJ....141...71F} and Table~\ref{tab:triples}.  By this metric, LP~704-48/{\namesh}AB stands out as the most internally stable low-mass triple known, slightly more stable than the NLTT~41135AB/NLTT~41136 system \citep{2010ApJ...718.1353I}.  With such a large semi-major axis ratio, we note that N-body simulations find the outer orbits of such systems are likely to have a significant eccentricity \citep{2002A&A...384.1030S}, so the ratio of outer periapse to inner apoapse may be smaller than the semimajor axis ratio used here.  

\subsection{On the Formation of LP~704-48/{\namesh}AB}

Given the now established existence of a 
handful of VLM triples like LP~704-48/{\namesh}AB,
it is worth examining whether such systems and their characteristics match 
the predictions of star and brown dwarf formation models. We emphasize
that current samples are far from statistically robust or complete, and detection biases may be significant.

Despite the low production rate of VLM multiples in general, 
current models do produce VLM higher-order systems.  A commonly-ascribed mechanism is dynamical scattering in the post-accretion phase of a young cluster \citep{2003A&A...400.1031S,2004MNRAS.351..617D,2004A&A...414..633G}. In their study of small N-body cluster dynamics, \citet{2003A&A...400.1031S} found that 11\% of their simulated systems formed triple and higher-order multiples, and over 40\% of the triples contained at least one brown dwarf component.  Most of these systems were widely separated and hierarchical, typically with the distant companion being of low mass (in systems containing two brown dwarfs, the distant companion was a brown dwarf in 90\% of cases).   However, all wide triples ($a \gtrsim$ 100~AU) with brown dwarf components had extremely low mass ratios ($q < 0.3$), whereas the systems listed in Table~\ref{tab:triples} are largely composed of near-equal mass components.  Moreover, this simulation proved unable to generate VLM binaries tighter than $\sim$1~AU, and the authors conclude that dynamical interactions alone cannot explain the existence of such systems (which at the time included only PPl~15).  Hence, it would appear
that dynamics alone is incapable of producing VLM multiples
like LP~704-48/{\namesh}AB, a conclusion that has been reached for more massive higher-order multiples as well \citep{2008MNRAS.389..925T}.

Accretion plays an important role in the formation of multiple systems, as a mechanism for angular momentum exchange and dissipation, and driving binaries to near-equal masses (e.g., \citealt{2002MNRAS.336..705B}). 
\citet{2012MNRAS.419.3115B} recently examined VLM multiplicity in 
radiative hydrodynamic simulations of molecular cloud collapse, which incorporates both gas accretion (with some radiative feedback) and dynamics, albeit over shorter periods than pure N-body simulations due to computational constraints.
The Bate study produced three low-mass triple systems (M$_{tot}$ $\lesssim$ 0.6~AU), all of which were hierarchical with either the high mass or middle mass component at wide separation.
While an exact clone to LP~704-48/{\namesh}AB was not generated,
one system composed of a 0.15~{\msun} plus 0.03~{\msun}, 14~AU inner binary with a 0.07~{\msun} wide companion at 194~AU is a reasonable analog, although the large inner eccentricity of this triple ($e$ = 0.8) may ultimately lead to its dissolution.  One fundamental shortcoming of this study is the $\sim$1~AU separation range limit on close interactions, again due to computational constraints.  
\citet{2009MNRAS.392..590B} examined VLM binary statistics at ten times higher resolution but over a shorter timeframe and employing slightly different gas physics.  
While they could successfully form 0.1--1.0~AU VLM binaries
in these simulations, no analysis was made of higher-order multiples.

The fragmentation of circumstellar disks around massive stars is another  
proposed mechanism for forming VLM stars and brown dwarfs
\citep{2008A&A...480..879S,2009MNRAS.392..413S}.  Unfortunately, the predicted binary rate for VLM systems
created in these environments is very low ($\sim$8\% for ejected systems)
and no low-mass triples have been created in simulations to date.  Furthermore, while the separation
distribution of simulated substellar pairs peaks in a range that is close to that inferred
for {\namesh}AB (0.3 $< a <$ 0.6~AU), these binaries tend to have very large eccentricities ($e > 0.7$), a feature
not seen in any of  the VLM spectroscopic binaries identified to date.  As such, disk fragmentation does not appear to be a viable mechanism for making triples like LP~704-48/{\namesh}AB.

It is important to emphasize that current models fall short in reproducing 
very short-period, low-eccentricity VLM pairs like {\namesh}AB and 2MASS~J0320$-$0446AB. 
Such systems are necessarily the product of dynamical and dissipative evolution, as opacity-limited fragmentation constrains the initial separations of self-gravitating masses to $\gtrsim$10~AU \citep{1969MNRAS.145..271L,1976MNRAS.176..367L}.
\citet{2002MNRAS.336..705B} resolves this problem through a combination of
dynamical interactions and accretion from a circumbinary/circumtertiary disk, although 
again these simulations fail to produce stable VLM multiples. 
Indeed, three-body encounters more often than not disrupt VLM multiples \citep{2010MNRAS.404..721M}.
A more intriguing mechanism for {\namesh}AB is Kozai-Lidov eccentricity perturbations induced by LP~704-48, followed by circularization through tidal friction (KCTF; \citealt{1962AJ.....67..591K,1968AJ.....73..190H,1998MNRAS.300..292K,2007ApJ...669.1298F,2012ApJ...750..106S}).  This mechanism has been proposed to explain the high fraction of spectroscopic binaries with tertiary companions (96\% for SBs with $P < $3~dy; \citealt{2006A&A...450..681T}; see also \citealt{2010ApJS..190....1R}).  KCTF predicts end states with roughly circular inner orbits and large period ratios ($X \gtrsim 10^3-10^4$; \citealt{2007ApJ...669.1298F}), similar to {\namesh}AB.  Unfortunately, the timescale for eccentricity pumping is long for LP~704-48/{\namesh}AB ($\tau {\sim} P^2_{out}/P_{in}$ $\approx$ 3~Gyr), while tidal circularization is extremely inefficient at the current orbital separation of {\namesh}AB\footnote{Using Eqns~50 and 52 from \citet{1981A&A....99..126H}, the timescale for circularization:
\begin{equation}
t_{circ} = \frac{0.0024}{kq}\left(\frac{a_0}{R_1}\right)^8\frac{P_0}{\tau}P_0
\end{equation}
where $k$ is the primary's Love number, $q$ the mass ratio, $a_0$ and $P_0$ the final semimajor axis and period, $R_1$ the primary's radius and $\tau$ the tidal lag with respect to the displacement between primary and secondary.  Assuming $k \sim 0.5$ (i.e., comparable to Jupiter and Saturn), $q = 0.7$, $a_0/R_1$ = 0.29~AU/$R_{Jupiter} \approx 620$ and $P_0$ = 0.4~yr and $\tau \approx 0.1P_0$, we derive $t_{circ} \approx 6{\times}10^{20}$~yr, far too long to have played a role in the dynamic evolution of {\namesh}AB.}  \citep{1981A&A....99..126H,2005ApJ...620..970M}.  Thus, it appears that KCTF is probably not responsible for the current configuration of this system, nor presumably for the isolated tight VLM binaries PPl~15AB, 2MASS~J0535$-$0546AB or 2MASS~J0320$-$0446AB.

A final possibility is that LP~704-48 has played no role in the evolution of the {\namesh}AB binary, which could have shrunk and circularized through dissipative interactions with an early circumbinary disk or a closer encounter resulting in an ejection.  The wider pairing with LP~704-48 may simply be a normal (albeit rare) outcome of cloud fragmentation or parallel trajectories during the dissolution of the natal cluster \citep{2010MNRAS.404.1835K,2010MNRAS.404..721M}.
LP~704-48 and {\namesh}AB may be more akin to stellar cousins that stellar siblings.

\section{Summary}

We have identified a hierarchical triple system, LP~704-48/{\namesh}AB,  composed of a tight ($a_{in} = 0.28$~AU) M9 plus T5$\pm$1 binary straddling the hydrogen burning mass limit, with a wide ($a_{out} \approx$ 820~AU), inactive M7 co-moving companion.  The inner pair is found to be both a spectral (blended light) binary and a radial velocity variable.  
By combining model-dependent mass constraints from the component spectral types, the radial velocity of LP~704-48, and orbit model fits of the RV data (spanning more than three orbital periods), we have determined the
orbital parameters of {\namesh}AB, including constraints on the orbit inclination and the system's age ($\gtrsim$3~Gyr). The latter 
is consistent with the age inferred for LP~704-48 based on its lack of H$\alpha$ emission (8$^{+0.5}_{-1.0}$~Gyr). {\namesh}AB is currently the third tightest VLM binary known, and its identification verifies that the spectral binary method can probe a relatively unexplored separation range among VLM multiples.
With LP~704-48, this system forms the lowest-mass triple identified to date whose tertiary is unambiguously detected, and is one of seven confirmed and candidate VLM triples whose total mass is $\lesssim$0.3~{\msun}.  While current star formation models are capable of producing triples with substellar components, we find that
they do not yet produce systems that replicate VLM triples like LP~704-48/{\namesh}AB.
In particular, dynamical interactions alone do not appear to be responsible for the close separation of the {\namesh}AB pair, while disk fragmentation and hydrodynamic models have yet to produce comparable systems.
We also rule out Kozai-Lidov perturbations and tidal circularization for the current configuration of this system.  
As most of the $P < 1$~yr VLM binaries found to date are isolated
pairs, we speculate that LP~704-48 and {\namesh}AB formed as two separate systems bound only by proximity and/or common motion after cluster dissolution.  Assessing whether this is the primary mode for VLM triple formation, and whether VLM triples are common or rare, will require a more statistically robust survey.  Nevertheless, these systems are benchmarks for studying multiple star formation theory; brown dwarf evolutionary models; and age, activity and metallicity diagnostics for VLM stars.

\acknowledgements
The authors thank
Joel Aycock, Scott Dahm, Heather Hershley, Carolyn Parker, Jim Lyke, Julie Rivera, and Greg Wirth at Keck Observatory;
Bobby Bus, Bill Golisch and John Rayner at IRTF; 
and Dave Summers at KPNO for their assistance with the observations.
The authors acknowledge helpful comments from Trent Dupuy and Tristan Guillot.
AJB acknowledges Rob and Terry Ryan for providing accommodation while completing this manuscript.
We also thank our referee, John Johnson, for his prompt and insightful review.

This publication makes use of data products from the Two Micron All Sky Survey, which is a joint project of the University of Massachusetts and the Infrared Processing and Analysis Center/California Institute of Technology, funded by the National Aeronautics and Space Administration and the National Science Foundation. 
This article has also made use of data from the Sloan Digital Sky Survey.    Funding for the SDSS and SDSS-II has been provided by the Alfred P. Sloan Foundation, the Participating Institutions, the National Science Foundation, the U.S. Department of Energy, the National Aeronautics and Space Administration, the Japanese Monbukagakusho, the Max Planck Society, and the Higher Education Funding Council for England. The SDSS Web Site is http://www.sdss.org/.
The SDSS is managed by the Astrophysical Research Consortium for the Participating Institutions. The Participating Institutions are the American Museum of Natural History, Astrophysical Institute Potsdam, University of Basel, University of Cambridge, Case Western Reserve University, University of Chicago, Drexel University, Fermilab, the Institute for Advanced Study, the Japan Participation Group, Johns Hopkins University, the Joint Institute for Nuclear Astrophysics, the Kavli Institute for Particle Astrophysics and Cosmology, the Korean Scientist Group, the Chinese Academy of Sciences (LAMOST), Los Alamos National Laboratory, the Max-Planck-Institute for Astronomy (MPIA), the Max-Planck-Institute for Astrophysics (MPA), New Mexico State University, Ohio State University, University of Pittsburgh, University of Portsmouth, Princeton University, the United States Naval Observatory, and the University of Washington.
This research has benefitted from the M, L, and T dwarf compendium housed at \url{http://DwarfArchives.org} and maintained by Chris Gelino, Davy Kirkpatrick, and Adam Burgasser;
the the Very-Low-Mass Binaries Archive housed at \url{http://www.vlmbinaries.org} and maintained by Nick Siegler, Chris Gelino, and Adam Burgasser;
and the SpeX Prism Spectral Libraries, maintained by Adam Burgasser at \url{http://www.browndwarfs.org/spexprism}.
This research has also made use of the SIMBAD database and VizieR service, operated at CDS, Strasbourg, France. 

The authors recognize and acknowledge the 
very significant cultural role and reverence that 
the summit of Mauna Kea has always had within the 
indigenous Hawaiian community.  We are most fortunate 
to have the opportunity to conduct observations from this mountain.

%\bibliographystyle{../../apj}
%\bibliography{../../biblibrary}

\end{document}